%% file: book.tex
\documentclass[11pt]{book}

\usepackage{epsfig}
\usepackage{amssymb}
\usepackage{amsmath}

\usepackage{index}

\makeindex

\begin{document}

\begin{titlepage}
  \begin{center}

    \vspace*{5cm}

  \Huge
  \hrule
  \bigskip
  \textbf{Progress in Dark Matter Research}
  \bigskip
  \hrule

  \vfill
  \Large
  Edited By xxxxxxxxx
  \bigskip

  \LARGE

  Editors   yyyyyyyyy

  \vfill
  \newpage
  \thispagestyle{empty}
  \verb+ +
  \end{center}
\end{titlepage}

\pagenumbering{roman}

\cleardoublepage 
\setcounter{page}{5}
\tableofcontents

\pagenumbering{arabic}

\include{fatdm}    

\input{fatdm.bbl}


\end{document}

%% file: fatdm.tex
\def\phiq{{\phi}_q}
\def\phix{{\phi}_x}

\def\etal{{\it et al.}}
\def\st{\scriptstyle}
\def\sst{\scriptscriptstyle}
\def\epp{\epsilon^{\prime}}
\def\vep{\varepsilon}
\def\ra{\rightarrow}
\def\ppg{\pi^+\pi^-\gamma}
\def\vp{{\bf p}}
\def\ko{K^0}
\def\kb{\bar{K^0}}
\def\al{\alpha}
\def\ab{\bar{\alpha}}
\def\be{\begin{equation}}
\def\ee{\end{equation}}
\def\bea{\begin{eqnarray}}
\def\eea{\end{eqnarray}}
\def\CPbar{\hbox{{\rm CP}\hskip-1.80em{/}}}

\def\st{\scriptstyle}
\def\sst{\scriptscriptstyle}
\def\epp{\epsilon^{\prime}}
\def\vep{\varepsilon}
\def\ra{\rightarrow}
\def\ppg{\pi^+\pi^-\gamma}
\def\vp{{\bf p}}
\def\ko{K^0}
\def\kb{\bar{K^0}}
\def\al{\alpha}
\def\ab{\bar{\alpha}}
\def\be{\begin{equation}}
\def\ee{\end{equation}}
\def\bea{\begin{eqnarray}}
\def\eea{\end{eqnarray}}
\def\CPbar{\hbox{{\rm CP}\hskip-1.80em{/}}}

\chapter{Quest for Fats: Roles for a Fat Dark Matter (WIMPZILLA)}

\begin{center}
{\it Houri Ziaeepour\\Mullard Space Science Laboratory\\Holmbury St. Mary, 
Dorking, RH5 6NT Surrey, UK.}
\end{center}

Since 1990's the detection of extremely energetic air showers and precise 
astronomical measurements have proved that our knowledge about fundamental 
laws of Nature is far from being complete. These observations have found 
convincing evidences against two popular believes: The spectrum of Cosmic 
Rays would have a steep cutoff at energies around $10^{19}eV$ (GZK cutoff) 
and the contravortial quantity called Cosmological Constant (dark energy) 
should be strictly zero. They have been important additions to the yet 
unsolved mystery of the nature of dark matter. 

For both phenomena many models have been suggested. The top-down model - 
decay of a Superheavy Dark Matter (SDM), also called WIMPZILLA as the 
origin of the Ultra High Energy Cosmic Rays (UHECRs) - is one of the most 
favorite candidates. Here we show that a meaningful constraints on the mass, 
lifetime and cosmological contribution of SDM is possible only if the energy 
dissipation of the remnants is precisely taken into account. We discuss the 
simulation of relevant processes and their evolution in the cosmological 
environment. We show that such a dark matter can be the dominant component 
of Cold Dark Matter (CDM) with a relatively short lifetime. Moreover, the 
equation of State of the Universe in this model fits the Supernova type Ia 
data better than a stable dark 
matter. If a small fraction of the mass of the SDM decays to an axion-like 
scalar field, its condensation can also explain the dark energy without 
need for extreme fine tuning of the parameters. Presumably, a meta-stable 
dark matter can explain 3 mysteries of Physics and Cosmology. Finally we 
review some of the particle physics and cosmological issues related to SDM 
and its associated quintessence field.

\section {Introduction} \label {sec:intro}
First evidences for the existence of a dark component of matter with only 
gravitational effects in cosmological environment (galaxy clusters) was 
discovered in 1930's~\cite{zwicky} and investigated more in detail with 
first measurement of galaxy-galaxy correlation during 1950's~\cite{galvelo}. 
It was however only after 1980's large galaxy surveys and measurement of 
Milky Way velocity curve~\cite{mwvelo} that its existence became an 
established fact. Roughly at the same time in Particle Physics, growing 
interest and efforts were dedicated to Super Symmetric (SUSY) 
models~\cite{susyinit}. Naturally many cosmologist and particle physicists 
found supersymmetric partners of ordinary matter and more specially the 
lightest of them (LSP) as the best candidate for the mysterious Dark 
Matter (DM). 

Generally, CDM is assumed to be composed of Weakly 
Interacting Massive Particles (WIMPs). Until today neither of efforts for 
direct detection of these particles nor searches for their signature in 
astronomical data have found any reliable signal. Search for SUSY 
particles including LSP in accelerators is also 
in the same situation. Nonetheless, next generation of high energy particle 
colliders like LHC have a good chance to detect supersymmetric particles 
if they exist and if their mass is less than few hundred GeV~\cite{lhcreach}. 

Constraints on the coupling of WIMPs to baryonic matter depend on the 
assumptions about their mass and their flux around the Solar system, the 
type of their interaction with baryonic matter (branching ratio, spin 
dependence, etc.) and their self-annihilation cross-section. These 
quantities are usually estimated based on the assumption that dark matter 
(WIMPs) are LSP. The reason for this apriori is that in the light of new 
experiments other potential 
candidates like left hand neutrinos and QCD axion are proved to have very 
small contribution to the total CDM~\cite{wmapres}~\cite{axionrev}.

Minimal Super Symmetry Model (MSSM) has more than 100 parameters and 
therefore should be considered not as a real model but as a framework. 
Even its constrained form (CMSSM)~\cite{cmssm}, depends on multiple 
parameters like Higgs multiplet mass parameter $\mu$, Higgs masses and 
ratio of Vacuum Expectation Values (VEVs) $\tan \beta$, gaugino mass 
$m_{1/2}$ and scalars' masse $m_0$ at unification scale, 
etc. Many but probably not all of these parameters can be determined by 
LHC if SUSY scale is not much higher than electroweak scale presumably 
$\sim 1 TeV$~\cite{lhcrev}. Part of the parameter space is already excluded 
by recent precise measurement of WMAP~\cite{wmapres}, limits on the mass of 
Higgs, muon anomalous magnetic momentum $g_{\mu} - 2$ and neutrino mixing and 
texture ~\cite{csmmsex}~\cite{lhcreach}. Nonetheless, large part of the 
parameter space is yet possible. In addition, relaxing some of the 
constraints 
considered as being {\it realistic} increases the allowed ranges.
Consequently the contribution of LSP in DM is not well determined 
and a large number of combinations are apriori permitted. For instance, the 
region with neutralino mass $m_{\chi} \gtrsim 1.4 TeV$ gives a contribution 
which contradicts WMAP observation and therefore is ruled out. By contrast 
many other combinations of parameters can lead to a contribution much 
smaller then observed ${\Omega}_{CDM}$.

LSP is usually considered to be a neutralino, i.e. the lightest gauginos 
which is assumed to be a mixture of bino and higgsino. Its mass can also be 
very close to stau mass. LSP must be stable or has a long lifetime if 
${\mathcal R}-$parity is conserved, otherwise most probably (but it depends 
again on the other parameters) it will have a short lifetime and can not 
contribute to the dark matter. Therefore even the detection of LSP in 
accelerators is not 
a proof that dark matter is LSP. One has to find its lifetime which is not 
very easy if it does not decay inside detectors. The only possibility 
in a near future to find such a signal is astronomical data. At present no 
evidence has been found, but as usual channels to search for such a 
signal as well as self-interaction of LSP depend on the unknown parameters 
of SUSY models.

Cosmological contribution of a stable or meta-stable particle - a decaying 
particle with a lifetime of order or longer than present age of the 
Universe - depends on its mass, the cross-section of its interaction with 
itself (self annihilation) and other species, its kinematic i.e. if 
it had sufficient interactions in the early Universe - presumably after 
inflation and reheating - such that its distribution becomes thermal. To 
make a rough 
estimation about the relation between mass and interaction cross-sections 
we can use a simple form of Boltzmann equation for cosmological evolution 
of $n (t)$ the number density of a stable species $X$ (thermal or chemical 
equilibrium is not assumed):
\be
\frac {dn}{dt} = - 3 (w_q + 1) H n - \langle \sigma v \rangle n^2 - 
\langle {\sigma}_{N} v \rangle n N + \langle {\sigma}_n v \rangle N^2\label {lspn}
\ee
where $w_q$ determines the equation of state of $X$, $\sigma$,  
${\sigma}_N$ and ${\sigma}_n$ are respectively total cross-section for 
its self-annihilation, its interaction with other particles and its 
production in the interaction of other particles collectively called $N$, 
$v$ is a nominal velocity 
determined by kinetic energy of particles, and $H (t) = \dot {a} / a$ is 
the Hubble Constant at time $t$. For weak interacting particle (as a DM 
should be) one expects that the second and third terms on the right hand 
side of 
(\ref {lspn}) must be less significant than others. Neglecting these terms, 
the decoupling of the particle happens when expansion term becomes dominant. 
If at this time the value of Hubble Constant is determined by other species, 
the only factor which determines today contribution of $X$ in DM is its 
self-annihilating cross-section and its mass. Higher masses need a larger 
cross-section (or lower initial density) to respect the observed mass 
constraint. Conversely, if $X$ is relatively light and its self-annihilation 
cross-section is relatively high, its contribution to CDM can be small. 
If the claimed deviation of $g_{\mu} - 2$ from predicted value by Standard 
Model is confirmed, this quantity put a more stringent lower limit on the 
self-annihilation cross-section than Higgs lower mass limit~\cite{csmmsex}. 
As this limit is much 
higher than the limit from ${\Omega}_{CDM}$, either LSP has only a small 
contribution in the CDM (roughly comparable to the contribution of baryonic 
matter) or its mass must be $m_{\chi} \gtrsim 200 GeV$, just in the limit of 
LHC observation possibility.

In summary it is worth to look for other potential components for CDM 
although it is not yet sure that in addition to LSP we need another stable 
or meta stable component. Above arguments are not in fact the only 
motivations for looking for other alternatives. Here we briefly review other 
reasons in favor of non-LSP candidates for dark matter and specially ones 
with masses around Grand Unification (GUT) scale.

In the theoretical side various 
phenomena at SUSY scale and beyond are able to make new hierarchy of masses 
and conserved quantum numbers. Many models can found in literature. Some 
examples are listed in ~\cite{gutmodels}\footnote{In particle 
physics for each of these issues large number of models have been 
investigated. As it is impossible to list all these works here, we give 
only a few examples of ideas and therefore references mentioned here as well 
as variety of ideas are not exclusive.}. 
For instance in gauge mediated SUSY breaking 
models one of the reasons for supersymmetry breaking in the hidden sector 
can be the condensation of gauginos which can also lead to gauge symmetry 
breaking (For a review of SUSY breaking mechanisms see ~\cite{susybreak}). 
If this creates a split in the masses in the hidden sector similar to 
$SU (2)-SU (3)$ splitting in SM sector, some of fields can get strong 
interactions. This phenomenon along with existence of conserved (or 
approximately conserved) global symmetries due to splitting can make very 
massive and long lifetime particles. Discrete symmetries can also make 
massive fields like messenger bosons meta-stable and therefore a good 
candidate for SDM~\cite {highmass}.

Since 1998 various observations including the observation of high redshift 
Supernovae Type Ia~\cite{snmeasur}~\cite{newsn}, precise CMB anisotropy 
observation by WMAP~\cite{wmapres}, large galaxy surveys like 
SDSS~\cite{sdss} and correlation between them~\cite{swsdss} as well as 
comparison between images of lensed distance objects~\cite{lensdm} show 
that the energy contents of the Universe is dominated 
by a mysterious form of energy called Dark Energy (DE) with an equation of 
state very close to a cosmological constant. In the following sections we 
will argue in detail that somehow there must be a relation between 
dark matter and dark energy. Some of the candidate theories which can 
provide solutions for mass hierarchy contain also axion like 
particles~\cite{axionmod} which can play the role of a Quintessence 
field~\cite{quin}.

On the experimental side the main motivation for considering scenarios in 
which part of the dark matter is a very heavy particle is the mystery of 
observed Ultra High Energy Cosmic Rays (UHECRs). Study of such possibility 
is the main subject of this chapter. We therefore leave the detail 
explanation to the next section and conclude this introduction by 
summarizing the contents of following sections. 

In this chapter we show that apriori it is possible to relate 
three problems we mentioned in above i.e. dark matter, dark energy and 
UHECRs. Our solution assumes the existence of a super 
heavy dark matter. We first obtain constraints on the properties of these 
particles like mass and lifetime. Then we study their effect on the 
cosmological equation of state. We also consider the role they can have in 
producing a light scalar field which is able to explain the dark energy and 
its equation of state. Finally we briefly discuss some of particles physics 
models with proper field contents and the issue of their production in the 
realy Universe.

\section {Ultra High Energy Cosmic Rays} \label {sec:uhecr}
Principal motivation for existence of an ultra heavy meta-stable particle 
has not been originally the quest for dark matter but the observation of 
Ultra High Energy Cosmic Rays (UHECRs) by large air shower 
detectors~\cite{yakutsk}~\cite{fly}~\cite{agasa}~\cite{sugar} (For a review 
of UHECRs detection and observed properties see ~\cite{crrev}). 
The predicted GZK cutoff~\cite{cmbir} in the spectrum of CRs at energies 
around $\gtrsim 10^{19} eV$ due to interaction with CMB and IR photons 
restricts the distance to the source to less than $\sim20-50 Mpc$, depending 
on the injection rate, spectrum, and on the amount of background radiations 
specially radio and IR. Optical depth of protons around GZK cutoff can be 
roughly estimated by $\tau_{opt} \approx \sigma n_{cmb}$. For 
$\sigma \approx 0.45 mb$~\cite{pphrev} close to $\pi$ production resonance, 
the probability of propagation without interaction in a distance of $30 Mpc$ 
is at most $\sim 10^{-8}$. 

Order of magnitude statistics of observed events is: $\sim 1000$ events with 
$E \gtrsim 10^{19} eV$, $\sim 100$ with $E \gtrsim 4 \times 10^{19} eV$ and 
20 events with $E > 10^{20} eV$~\cite{eveclust}~\cite{evelist} including 
one with $E \sim 10^{21} eV$~\cite{fly}. The UHECRs spectrum roughly has the 
same power-law shape up to around $E \sim 10^{19} eV$. But at higher 
energies spectrum becomes flatter in contrast to prediction of GZK cutoff.

Composition of the primary particles~\cite {compo}~\cite {comp2} can be 
estimated from the shower, specially from maximum position and elongation 
rate of muons in the atmosphere. Uncertainties in determination of primaries 
composition include dependence on the hadrons interaction/fragmentation model 
at high energies~\cite{hadmod} and on the detector response. However, most 
analyses are in favor of a hadronic particles $p, \bar {p}, n, \bar {n}$. 
Light nuclei like $D$ or $He$ can not be completely ruled 
out~\cite{fly}~\cite{compo}~\cite{comp2}. 
This composition is very different from one at lower energies which is 
dominated by $Fe$ nuclei and is probably an evidence of a different 
origin for Cosmic Rays with $E \gtrsim 10^{19} eV$. Some authors have tried 
to explain this modification of composition by disintegration of Iron and 
other heavy nuclei in the cosmic photon field. By considering a unique 
injection energy of $10^{22} eV$ for $Fe$ (i.e. $\sim 10^{20} eV$ per 
nucleon)~\cite{nucldisinteg} a roughly constant distribution of nucleons 
up to a distance of $50 Mpc$ from the source has been found. It has been 
claimed 
that up to simulation precision this result does not strongly depend on the 
extra-galactic magnetic field~\cite{nucldisinteg}. By 
contrast, the latter affects the flux of Iron and other nuclei. We will 
see later that high energy nucleons lose large amount of energy to CMB and IR 
background 
during propagation and a constant distribution does not seem realistic. In 
fact another similar study~\cite{nucldisinteg0} finds 
that when the injection energy is limited to $Z \times 2 \times 10^{19} eV$, 
at the same distance of $\sim 50 Mpc$, protons are concentrated in 
$E \lesssim 10^{19} eV$ and 
therefore it seems that it is difficult to explain the composition change by 
disintegration. Correlation between Super Galactic Plane and clustering in 2 
doublets and one triplet was claimed~\cite {eveclust}~\cite{corr}~\cite
{evelist}, but denied by other analyses~\cite{nocorr}~\cite{noclust}. 
Moreover, apparent clustering of events can originate from caustics 
generated by the galactic magnetic field~\cite {crprog}~\cite {magdev}

\subsection {Origin of UHECRs} \label {sec:origcr}
Few phenomena in the history of physics have had as many suggested origins 
as UHECRs. From the 
most classical sources (if the word {\it classic} makes any sense when we 
talk about the most extreme objects and environments we can find today in 
the Universe) i.e. shocks in the supernovae remnants or somehow more exotic 
astronomical objects like accretion disk around supermassive black holes in 
Active Galactic Nuclei (AGNs) or dormant AGNs like one in the center of our 
own galaxy, up to modifications in some of the most fundamental laws of 
physics like Lorentz invariance of the space-time have been proposed as the 
origin og UHECRs.

Here we discuss some of conventional and exotic potential sources of UHECRs 
very briefly. Our main purpose is to show that a SDM i.e. top-down solution 
is at present one of the most plausible candidate until Auger Observatory 
solves issues like composition of primaries, anisotropy and its relation 
with galactic halo and local matter over-densities (like Virgo Cluster), 
ultimate break in the high energy tail of the spectrum, etc.

\subsubsection {Conventional Candidates}\label {sec:origcrconv}
Conventional accelerators can hardly accelerate protons to energies 
requested for UHECRs. Maximum energy a charged particle can be accelerated to 
by Fermi mechanism~\cite{fermiacc} is\footnote{Through this chapter we use 
unit system in which $c$ = $\hbar$ = 1}:
\be
E_{max} = \biggl (\frac {3 \eta B R^2}{2 eZ}\biggr )^{\frac {1}{4}} m \label{maxacc}
\ee
where $B$ is the magnetic field, $R$ is the size of acceleration zone 
($R \lesssim r_{Larmor} = E / eB$), and $\eta B$ is the effective electric 
field in the direction of particles trajectory. In this formula the only 
source of energy loss is considered to be synchrotron radiation. Using 
(\ref {maxacc}) and approximate knowledge about size and magnetic field of 
astronomical objects Table \ref{tab:fermiacc} shows maximum energy obtainable 
by Fermi acceleration in some of astronomical objects proposed as the 
source of UHECRs. Particles energy after escaping from acceleration zone is 
certainly smaller than what is shown in this table partly because of 
adiabatic deceleration when the magnetic field 
becomes small and partly because of high probability of interaction with 
particles in the acceleration zone or its outskirts~\cite{enerloss}. 
The suggested solution is the change of particle type from proton to 
neutron which then can escape the adiabatic deceleration~\cite{exnu}. 
But this needs interaction with other particles i.e. loss of significant 
amount of energy~\cite{enerloss}. Adiabatic expansion problem can also be 
solved by an abrupt change in the magnetic field, but this needs a fine 
tuning of the source structure.

\begin{table}[t]
{\caption{Maximum acceleration energy for protons by Fermi Mechanism}
 \label{tab:fermiacc}}
\vspace{0.2cm}
\begin{center}
\footnotesize
\begin{tabular}{p{25mm}|l|l|l|p{35mm}}
Object & Size & $B (G)$ & $E_{max} ()$ &  \\ \hline
Supernova Remnant~\cite{snhigh} & $1 pc$ & $\sim 10^{-5}-10^{-5}$ & 
$\sim 10^6-10^7 $ & Too small. \\ \hline
Close to central black hole in AGNs~\cite{accagn} / 
Quasar Remnants and Dormant AGNs~\cite{qsorem} & $R \sim 0.1 pc$ & 
$\sim 30$ & $\sim 10^{12}-10^{13} $ & Too far (see also the text). \\ \hline
Shock front of relativistic jets & $\sim 10^{-3} pc$ & $\sim 5$ & 
$\sim 10^{10}-10^{11} $ & Too small for events $E \gtrsim 10^{20} eV$.\\ \hline
Hot Spots, radio galaxies & $\sim 1-10 pc$ & $\sim 10^{-4}$ & $\sim 10^{11}-10^{12} $ & Too far. \\ \hline
GRBs~\cite{grb} & $\lesssim 0.1 pc$ & $\sim 10^{4}$ & $\sim 10^{10}-10^{11} $ & Energy and unconfirmed assumptions~\cite{multifron}.\\ \hline
Pulsars (Magnetars)~\cite {magnetar} & $\sim$ Few km & $\sim 10^{14}-10^{15}$ & $\sim 10^{12}-10^{13} $ & Too rare.
\end{tabular}
\end{center}
\end{table}

Other acceleration mechanisms like Alfven waves~\cite{alfven} and 
multi-front shocks~\cite{multifron} which can produce somehow higher 
maximum acceleration energies have been also proposed. However they 
don't solve the problem of energy loss completely. In addition, one of 
the most important constraints on acceleration models is the unobserved 
excess of TeV $\gamma$-rays~\cite{enerloss} and 
high energy neutrinos (although physics of high energy neutrinos at 
present is too uncertain to use them as a constraint). In the case of GRBs, 
a simulation~\cite{gphdel} of cosmological distribution of sources with 
a power-law flux of UHECRs shows that the expected flux on Earth 
is much lower than observed values.

\subsubsection {Correlation with Astronomical Objects} \label 
{sec:origcrcorr}
Many efforts have been dedicated to correlate UHECRs to astronomical 
objects, but practically all of them have been failed or were not confirmed 
by further investigations. The deflection in galactic magnetic field can be 
the reason if sources of UHECRs are extra-galactic objects. Our ignorance of 
the geometry and strength of magnetic in the galaxy and also in the local 
group does not permit a correct reconstruction of original direction of 
the primaries. 

Between all the correlation attempts there is probably one with interesting 
results because it does not try to correlate events one-by-one, but 
considers the global rate of correlation between UHECRs and nearby elliptical 
galaxies~\cite{ellipcorr}. Most of AGNs are in this type of galaxies. As 
the lifetime of AGNs 
is short - of order of $10^7$ years - one can suppose that dormant AGNs are 
in the center of these galaxies and the magnetic field of a Kerr black hole 
in their center can accelerate particles to very high energies as is 
suggested in ~\cite{qsorem}. Surprisingly, the maximum of correlation (less 
than the expected deflection angle by magnetic field) is not with center 
of these galaxies but at $\sim 1.5 R$ radius where one expects to be 
dominated by dark matter halo of the galaxy. Therefore the idea comes to 
mind that if there is a correlation, is it with the central AGNs or 
with dark matter halos around the galaxies ? Unfortunately the present data 
is too scarce to permit any meaningful conclusion.

Close to uniform distribution of UHECRs events has been concluded to be 
the evidence that UHECRs originate from some extra-galactic astronomical 
objects and not from a decaying SDM in the Galactic Halo~\cite{hanti}. 
MACHOs observation~\cite{machobs} however shows that inner Halo has a 
heterogeneous composition and a precise modeling of the anisotropy must take 
into account the distribution of various components as well as the shape  
of the halo and nearby high densities~\cite{aniso}. 

The effect of magnetic field is also very important both in determination of 
composition and in the estimation of the amount of anisotropy. We have 
already mentioned that doublet and triplet clustering can be due to the 
magnetic field. Simulations show also that depending on the composition of 
the primaries, the anisotropy can be amplified (for heavy nuclei) or smeared 
(for protons)~\cite{maganiso}. In addition, detailed 
investigation of present data shows that if AGASA and SUGAR estimation of 
flux is correct, potential extra-galactic nearby sources are not enough 
and a top-down source is needed~\cite{anisotopdown}. 

\subsubsection {Exotic Candidates} \label {sec:origcrexo}
List of non-classical (exotic) candidates and phenomena is much longer !
Acceleration in conventional sources is limited to existence of a plasma 
and presence of a magnetic field. In exotic sources (or non-sources !) 
various microscopic or macroscopic processes can be the acceleration cause.

Evaporation of primordial black holes (PBH) is one of the suggested 
sources. The Hawking temperature at the end of black hole life is 
enough high to produce extremely energetic elementary particles like 
quarks and gluons and thus UHECRs~\cite{pbh}. Production models of 
PBH~\cite{pbhspec} however need fine-tuning and PBH evaporation effect on 
reionization constrains their present number density~\cite{pbhnum}.

Topological defects were also a favorite potential source of 
UHECRs~\cite{defect}. But following CMB anisotropy observations which have 
not found significant contribution from defects in the CMB power spectrum, 
the interest on them as the source of UHECRs is fading.

Many model makers have tried to overcome the constraint on the distance of 
charge particles as UHECR primaries by either assuming that they are 
neutrinos or by considering that the primary nucleons are produced by 
neutrino 
interaction in the galactic halo or nearby over-densities. Modification of 
neutrino-nucleon cross-section in models with large 
extra-dimensions~\cite{extdim} or extensions to SM make neutrinos as 
possible primaries. These models increase neutrino-nucleon cross-section 
at high energies such that they can interact with terrestrial atmosphere 
and make UHECR showers. One problem with this class of models is 
that even if the cross-section of neutrino interaction becomes larger such 
that together with larger flux they can provide the observed flux of cosmic 
rays, one has to fine-tune the process such that it imitates a 
nucleon-nucleon interaction. 
The other problem specially regarding extra-dimensions is that UHECRs 
themselves strongly constrains the existence of low scale gravity and 
macroscopic extra-dimensions~\cite{extdimcons}. In addition, even in the 
case of existence of macroscopic extra-dimensions the smallness of 
interaction between bulk modes and fields on the visible brane prevents them 
to provide the observed flux of UHECRs~\cite{extdiminsuff}.

Probably the most popular model with ultra high energy neutrinos is Z-burst 
model~\cite{neuthalo}. In this model it is assumed that neutrinos with 
energies $\sim 10^{23} eV$ collide with a halo of neutrinos around Milky Way 
galaxy and create a hadronic jet which is observed as UHECRs. This model has 
various problems. Firstly it is very difficult to find a {\it conventional} 
source to accelerate charged particles to energies of order 
$\sim 10^{23} eV$ which then they can make ultra high energy neutrinos. 
The second blow to this model is the upper limit on the mass of the 
neutrinos from WMAP~\cite{wmapres} 
and neutrino oscillation experiments. In fact with a mass $\lesssim 0.2 eV$, 
it has been shown that the amount of over-density of neutrinos around a halo 
of $M \sim 10^{12}M_{\bigodot}$ is $\lesssim 2$~\cite{neuthalo1}. Although 
there are claims that for masses as low as $0.03 eV$ this process can 
provide the observed flux, there are strong constraints on these models from 
amount of low energy $\gamma$-ray background they should make during 
interaction with presumed neutrino hale~\cite{againtnu}.

Violation of Lorentz invariance~\cite{lorentzviol} due to quantum gravity 
effects which has been proposed for solving the puzzle of UHECRs is 
probably one of the most exotic suggestions. However, recent upper limits  
on the amount of Lorentz symmetry violation imposes strong constraints 
on this model~\cite{lorentzcons}.

Finally annihilation or decay of a superheavy dark matter particle is one 
the most popular non-conventional solutions. Existence of this class of 
particles was suggested long before the observation of 
UHECRs~\cite{crypton1}, but the quest of an explanation for unobserved GZK 
cutoff have given much more interest to search for particle models with 
necessary mass and lifetime. We first study the phenomenology of their decay 
and find constraints on their properties and other cosmological role they 
can play. Then in the last section of this chapter we briefly review some of 
suggested particle and production models.

\section {Decay / Annihilation of SDM} \label{sec:decay}
To be the source of UHECRs, SDM - sometimes called $X$ particles for 
simplicity - can be either meta stable with very small 
self-coupling~\cite{xpart} or self-annihilating stable 
particles~\cite{annih}. 
Their interaction with other species is considered to be very small and 
negligible. According to (\ref{lspn}) for self-annihilating case the density 
variation is proportional to square of density. Therefore annihilation 
cross-section must be enough small such that a sufficient number of them 
survive the high density of the early universe or their density at that time 
must be very high. By contrast at present the annihilation cross-section 
must be enough large to explain the flux of UHECRs. These constraints and 
conditions which go to opposite directions impose some fine-tuning on SDM 
particle physics model and 
characteristics. In the case of decay however all the requests go to the 
same direction. The interaction of $X$ with other particles must be very 
small. This decreases the probability of its production during 
preheating/reheating~\cite{reso} and increases its lifetime. Therefore, we 
only have 
to find the right lifetime such that its decay corresponds to the observed 
flux of UHECRs. Here we only study the decay of SDM. Nonetheless, 
most of our results can also be applied to annihilation. One simply 
needs to find the corresponding annihilation cross-section (or rate) 
for a given lifetime.

Decay/annihilation of SDM have important implications for the evolution of 
high energy backgrounds. This can be used to constrain the mass and lifetime 
of these particles. A realistic estimation of these parameters however 
is not possible without considering in detail the energy dissipation 
of decay/annihilation remnants. If this reprocessing of high energy 
particles is not taken into account, the lifetime/annihilation rate must be 
many orders of magnitudes larger than the age of the 
Universe~\cite {xx}~\cite{sarkar}~\cite{topdown}. A more complete 
simulation of decay and dissipation~\cite{houri} however shows that for 
the same mass range, the lifetime must be much shorter to explain the 
observed flux of UHECRs. This reduces the 
fine-tuning in the production mechanism. As we noted earlier, larger 
lifetime needs less interaction with other species and therefore it would 
be more difficult to make significant amount of $X$ particle in the early 
universe.

\subsection {Decay and Energy Dissipation} \label{sec:dissip}
There is no theoretically rigorous motivation which restrict the possible 
mass range of SDM particles. 
The only constraint is that if their decay remnant should explain UHECRs, 
their mass can not be less than $\sim 10^{21} eV$. The absolute upper limit 
according to present believes in particle physics is the Planck mass i.e. 
$\sim 10^{28} eV$. Assuming that $X$ particles must be related to physics at 
GUT scale i.e. $\sim 10^{25} eV$ rather than Planck scale, we can reduce 
the upper limit to this 
value. The suggested range for the lifetime according to theories is also 
very large $\tau_{X} \sim 10^{-3}-10^{10} \tau$ where $\tau_{X}$ and $\tau$ 
are respectively the lifetime of $X$ particles and the present age of the 
Universe~\cite{highmass}~\cite{crypton2}. 

The decay or annihilation modes of SDM depends on their unknown particle 
physics model. If we assume that their self-annihilation probability is 
small, they can not have electric charge or any other charge with relatively 
strong 
coupling. It is also likely that they don't decay directly to SM particles 
and their decay has a number of intermediate unstable states which decay 
in their turn. It is also very probable that remnants include stable WIMPs 
which are not easily observable~\cite{susydecay}. To study the maximal 
effects of remnants on high energy backgrounds, we assume that at the end, 
the whole decayed energy goes to stable SM particles. Therefore the best 
guess for their decay is to assume that it looks like hadronic decay 
channel of a heavy neutral 
particle i.e. $Z^{\circ}$ boson. Evidently it is not sure that they are 
bosons. But due to huge number of final particles after hadronization of 
primary partons, this does not significantly affect the decay remnants. 
To mimic the softening of energy spectrum due to multiple decay level, we 
assume that decay is similar to hadronization of a pair of gluon jets. 

For simulating the fragmentation and hadronization of initial gluons we use 
PYTHIA program~\cite{pythia}. It can not however properly simulate ultra 
high energy events, not only because of unknown physics at $10^{16} GeV$ 
scale, 
but also because of programming limits and it is necessary to extrapolate 
simulation results for $E_{CM} \leqslant 10^{20} eV$ up to 
$E_{CM} = 10^{24} eV$. In the simulation all particles except 
$e^{\pm}, p^{\pm}, \nu, \bar {\nu}$ and $\gamma$ decay. This is a valid 
assumption when particles propagate in cosmological distances. 
We neglect neutrino mass and for simplicity we don't distinguish neutrino 
flavors in the final fragmentation. 
Contribution of stable species in the total multiplicity and the total 
decay energy is summarized in Table \ref {tab:mult}. It shows that the 
mass of SDM has little effect on the composition of remnants. 

\begin{table}[t]
\caption{Energy and multiplicity contribution in remnants of SDM.
 \label{tab:mult}}
\vspace{0.2cm}
\begin{center}
\footnotesize
\begin{tabular}{|c|c|c|c|c|}
\hline
 & \multicolumn {2}{c|}{$M_{X} = 10^{24} eV$} & 
\multicolumn {2}{c|}{$M_{X} = 10^{22} eV$} \\
\hline
\raisebox{0pt}[13pt][7pt]{Part.} &
\raisebox{0pt}[13pt][7pt]{Ener. \%} &
\raisebox{0pt}[13pt][7pt]{Multi. \%} &
\raisebox{0pt}[13pt][7pt]{Ener. \%} &
\raisebox{0pt}[13pt][7pt]{Multi. \%}\\
\hline
\raisebox{0pt}[12pt][6pt]{$e^{\pm}$} & \raisebox{0pt}[12pt][6pt]{$6.7 \times 2$}
 & \raisebox{0pt}[12pt][6pt]{$9.7 \times 2$} & \raisebox{0pt}[12pt][6pt]{$6.7 \times 2$}
 & \raisebox{0pt}[12pt][6pt]{$9.8 \times 2$}\\
\raisebox{0pt}[12pt][6pt]{$p^{\pm}$} & \raisebox{0pt}[12pt][6pt]{$11.8 \times 2$}
 & \raisebox{0pt}[12pt][6pt]{$1.4 \times 2$} & \raisebox{0pt}[12pt][6pt]{$11.9 \times 2$}
 & \raisebox{0pt}[12pt][6pt]{$1.4 \times 2$}\\
\raisebox{0pt}[12pt][6pt]{${\nu} \& \bar{\nu}$} & \raisebox{0pt}[12pt]
[6pt]{$18.4 \times 2$} & \raisebox{0pt}[12pt][6pt]{$28.3 \times 2$}& \raisebox{0pt}[12pt][6pt]
{$18.1 \times 2$} & \raisebox{0pt}[12pt][6pt]{$28.3 \times 2$}\\
\raisebox{0pt}[12pt][6pt]{$\gamma$} & \raisebox{0pt}[12pt][6pt]{$26.2$}
 & \raisebox{0pt}[12pt][6pt]{$21$}& \raisebox{0pt}[12pt][6pt]{$26.6$}
 & \raisebox{0pt}[12pt][6pt]{$21$}\\
\hline
\end{tabular}
\end{center}
\end{table}

PYTHIA has been also used for determining the cross-section of interaction 
between remnants and the rest of baryonic contents of the Universe as well as 
CMB and other radiation backgrounds. For low energies ($E \lesssim 4 GeV$) 
where PYTHIA does not work properly and for elastic scatterings measured or 
analytical calculation have been used. Details can be found in ~\cite{houri}.

\section {Cosmological Evolution} \label{sec:evol}
To obtain constraints on the mass and lifetime of SDM, one should determine 
the expected flux of high energy remnants of its decay on Earth and 
compare them with observations. However, the lifetime and contribution of SDM 
particles in CDM are degenerate and one can increase one and decrease the 
other. For simplicity we assume that non-baryonic CDM consists only of SDM. 
When all other parameters (decay and fragmentation, etc.) are the same, 
this assumption leads to an upper limit on the lifetime of SDM. 

Boltzmann equations for space-time and energy-momentum distribution of 
particles are~\cite{ehler}:
\bea
p^{\mu}{\partial}_{\mu} f^{(i)}(x,p) - ({\Gamma}^{\mu}_{\nu\rho} p^{\nu} 
p^{\rho} - e_i F^{\mu}_{\nu} p^{\nu}) \frac {\partial f^{(i)}}{\partial 
p^{\mu}} & = & -({\mathcal A}(x,p) + {\mathcal B}(x,p)) f^{(i)}(x,p) + \nonumber \\
& & {\mathcal C}(x,p) + {\mathcal D}(x,p) + {\mathcal E}(x,p). \label {bolt}
\eea
\bea
{\mathcal A}(x,p) & = & {\Gamma}_i {p_i}^{\mu} {u_i}_{\mu} \quad\quad 
{u_i}_{\mu} \equiv \frac {{p_i}^{\mu}}{m_i}.\label {idec}\\
{\mathcal B}(x,p) &= & \sum_j \frac {1}{(2\pi)^3 g_i} \int d\bar p_j f^{(j)}(x,p_j) A (s){\sigma}_{ij}(s).\label {absint}\\
{\mathcal C}(x,p) & = & \sum_j {\Gamma}_j {p_j}^{\mu} {u_j}_{\mu} \frac {1}{(2\pi)^3 g_i} \int d\bar p_j f^{(j)}(x,p_j)\frac {d{{\mathcal M}^{(i)}}_j}{d\bar p}.\label {jdec}\\
{\mathcal D}(x,p) & = & \sum_{j,k}\frac {1}{(2\pi)^6 g_i}\int d\bar p_j d\bar p_k f^{(j)}(x,p_j)f^{(k)}(x,p_k) A (s) \frac {d{\sigma}_{j+k \rightarrow i+\ldots}}{d\bar p}.
\label {proint}
\eea
$x$ and $p$ are coordinates and momentum 4-vectors; $f^{(i)}(x,p)$ is the 
distribution of species $i$; $m_i$, $e_i$ and $\Gamma_i$, are its mass, 
electric charge and width $= 1/\tau_i$, $\tau_i$ is its lifetime; 
${\sigma}_{ij}$ is the total interaction cross-section of species $i$ with 
species $j$ at fixed $s$. Expression:
\be
\frac {d{\sigma}_{j+k \rightarrow i+\ldots}}
{d\bar p} = \frac {(2\pi)^3 E d\sigma}{g_i p^2 dp d\Omega} \label{sigdef}
\ee
is the Lorentz invariant differential cross-section of production of $i$ in 
the interaction of $j$ and $k$; $g_i$ is the number of internal degrees of 
freedom (e.g. spin, color); $d\bar p = \frac {d^3p}{E}$. We treat 
interactions classically, i.e. we 
consider only two-body interactions and we neglect the interference between 
outgoing particles. It is a good approximation when the plasma is not 
degenerate. It is assumed that cross-sections include summation 
over internal degrees of freedom like spin; 
$\frac {d{{\mathcal M}^{(i)}}_{j}}{d\bar p}$ is the differential 
multiplicity of species $i$ in the decay of $j$; 
${\Gamma}^{\mu}_{\nu\rho}$ is connection; 
$F^{\mu}_{\nu}$ an external electromagnetic field; and finally 
${\mathcal E}(x,p)$ presents all other external sources. $A (s)$ is a 
kinematic factor:
\be
A (p_i,p_j) = ((p_i.p_j)^2 - m_i^2m_j^2)^{\frac {1}{2}} = 
\frac {1}{2} ((s - m_i^2 - m_j^2)^2 - 4 m_i^2m_j^2)^{\frac {1}{2}}.
\label {kin}\\
\ee
where $A\sigma$ presents the probability of an interaction.

\begin{figure}[t]
\begin{center}
\psfig{figure=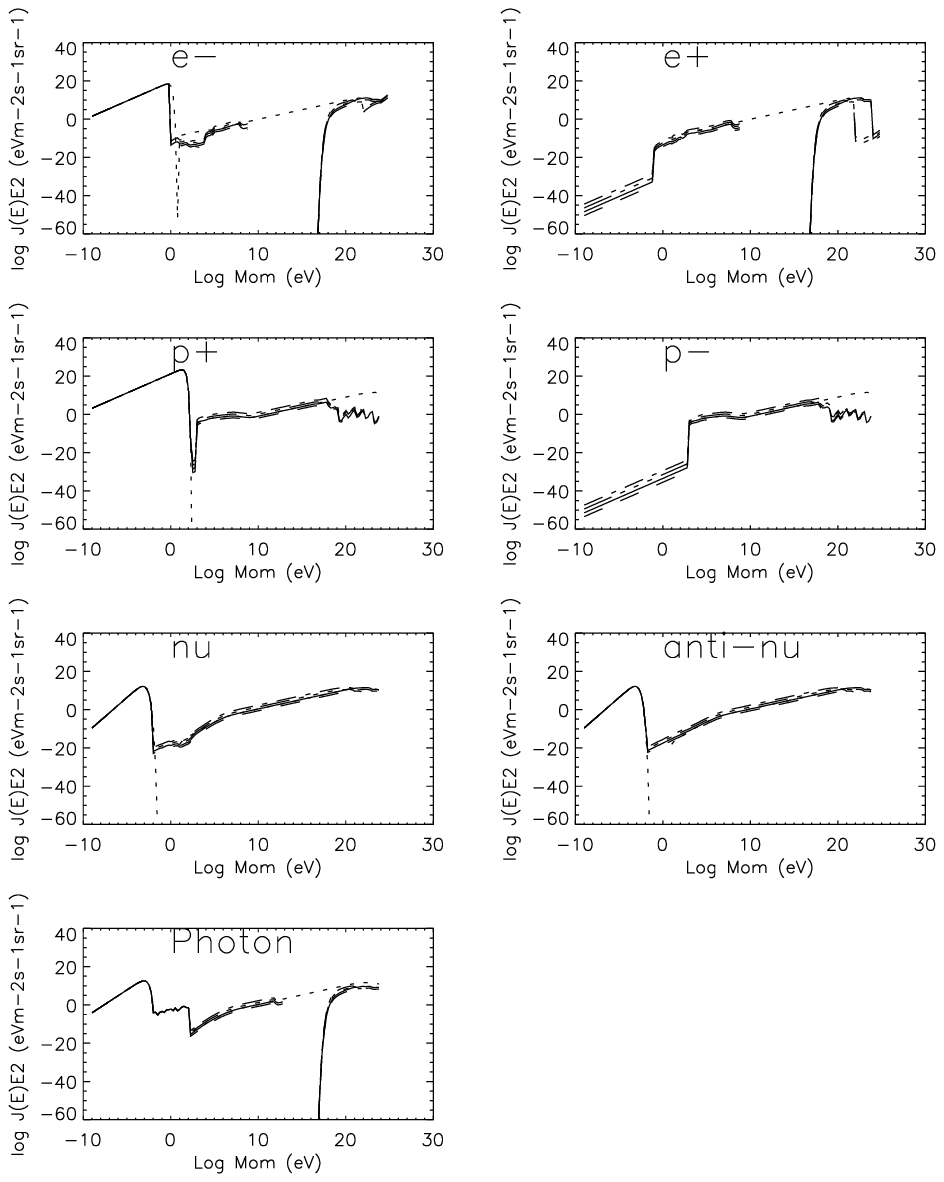,height=15cm}
\caption{Energy flux of stable species. 
Solid line 
$m_{dm} = 10^{24} eV$, $\tau = 5 \tau_0$, dot line is the spectrum 
without energy dissipation for the same mass and lifetime, dashed line 
$m_{dm} = 10^{24}eV$, $\tau = 50 \tau_0$, dash dot $m_{dm} = 10^{22} eV$, 
$\tau = 5 \tau_0$, dash dot dot dot $m_{dm} = 10^{22} eV$, 
$\tau = 50 \tau_0$.
\label{fig:distall}}
\end{center}
\end{figure}

\begin{figure}[t]
\begin{center}
\psfig{figure=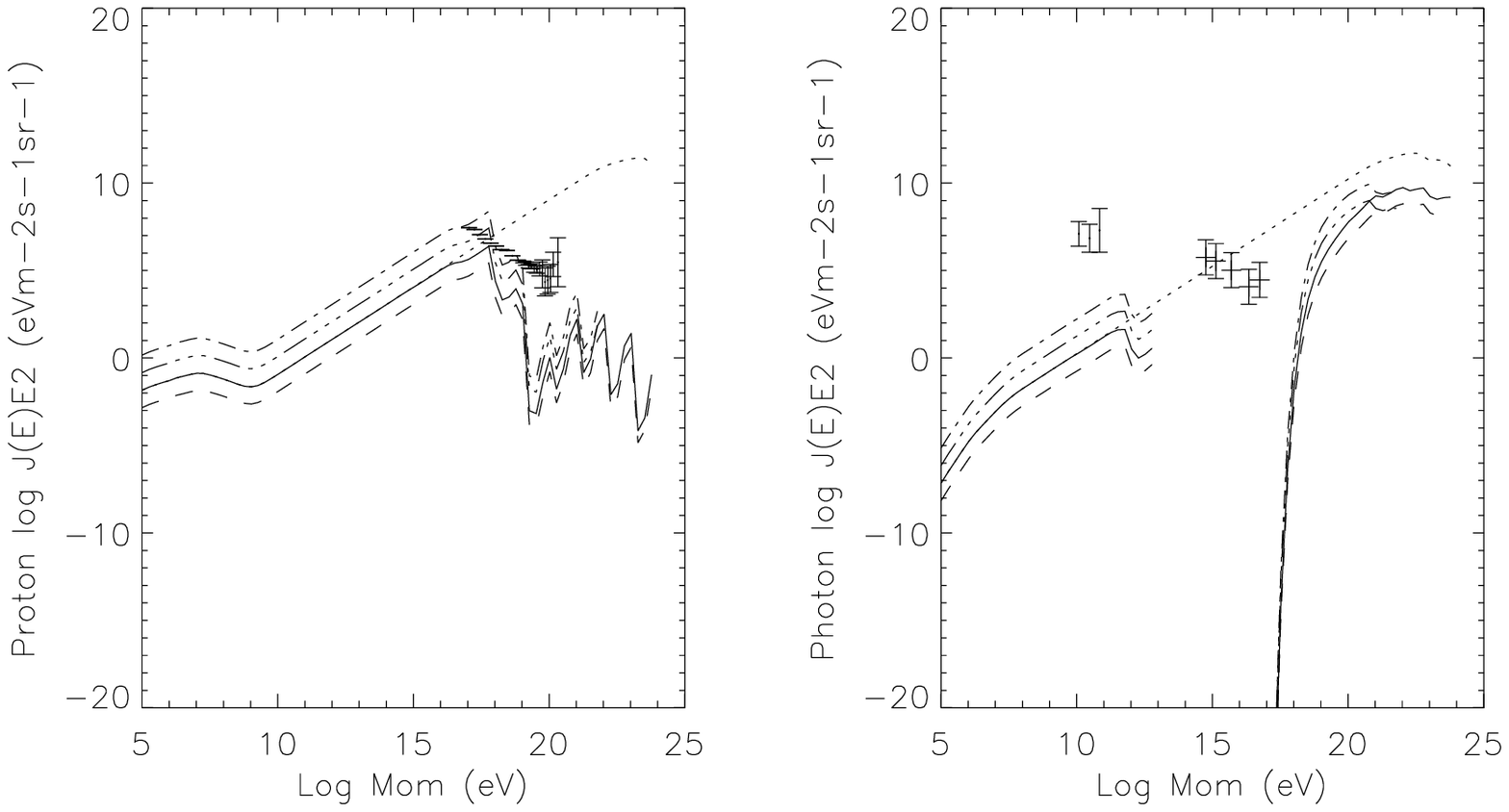,height=6cm}
\caption{Energy flux for protons and photons. Solid line 
$m_{dm} = 10^{24} eV$, $\tau = 5 \tau_0$, dot line is the spectrum 
without energy dissipation for the same mass and lifetime, dashed line 
$m_{dm} = 10^{24}eV$, $\tau = 50 \tau_0$, dash dot $m_{dm} = 10^{22} eV$, 
$\tau = 5 \tau_0$, dash dot dot dot $m_{dm} = 10^{22} eV$, 
$\tau = 50 \tau_0$. For protons, data from Air Showers 
detectors~\cite{crrev} is shown. Data for photons are EGRET whole sky 
background~\cite{egret} and upper limit from 
CASA-MIA~\cite{casa}.\label{fig:pgzoom}}\end{center}
\end{figure}

Using cross-sections discussed in Sec.\ref{sec:dissip}, this system of 
equations along with Einstein equation can be solved numerically. The flux of 
high energy stable species for a homogeneous cosmology is shown in 
Fig.\ref{fig:distall} and in Fig.\ref{fig:pgzoom} the calculated flux for 
protons and photons has been compared with observations. The interesting 
conclusion (which was first noticed in ~\cite{xx}) but for a much longer 
lifetime) is that even for a relatively short lifetime of $5\tau-50\tau$ 
for which we have simulated the decay process, the observed flux of UHECRs 
is orders of magnitude higher than what a uniform distribution of SDM 
can provide. Therefore it is necessary to consider the effect of clumping of
SDM in the galactic halo.

\begin{figure}[t]
\begin{center}
\psfig{figure=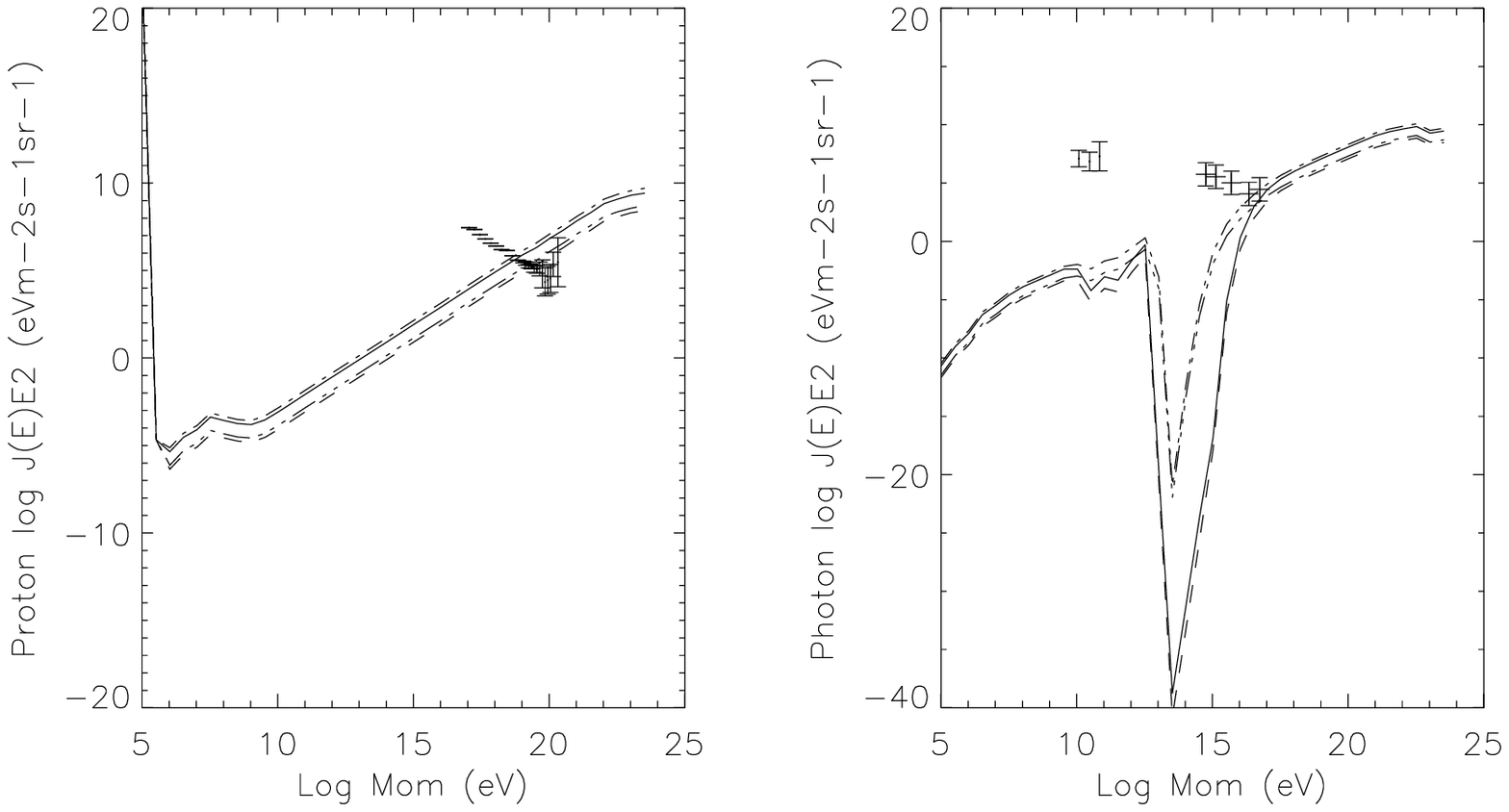,height=6cm}
\caption{Flux of high energy protons and photons in a uniform clump. $m_{dm} 
= 10^{24} eV$, $\tau = 5 \tau_0$ and $\tau = 50 \tau_0$. Dash dot and dash 
dot dot dot lines presents SDM halo. Solid and dashed lines show a halo of 
SDM 
and MACHOs. Data is the same as in Fig.\ref{fig:pgzoom}. For protons the 
effect of increasing lifetime of SDM is more important than presence of 
MACHOs. Photons trough is more sensitive to presence of MACHOs.
\label{fig:machhalo}}
\end{center}
\end{figure}

The simulation of halos even if we consider them to 
have a spherical symmetry is much more complicated than a homogeneous 
distribution. To simplify the task we simply consider a spatially limited 
halo with average over-density of $\delta = 200$ i.e. equivalent to the 
over-density at virial radius. For a halo of mass 
$M_{H} = 6 \times 10^{12} M_\odot$ the size of the virial radius is 
$r_{200} \sim 120 kpc$ (for NFW profile~\cite{nfw}). We assume a total size 
of $300 kpc$. We consider two distribution for the baryonic matter. In the 
first case there is 
no segregation between baryonic and non-baryonic matter. However the result 
of MACHOs observations show that it is possible that the inner 20 kpc of the 
halo is dominated by baryonic dark matter. Therefore in the second case we 
consider such a situation. Results for protons and photons is shown in 
Fig.\ref{fig:machhalo}.

According to this plot, the effect of clumping is more significant for 
protons than for photon which are more sensitive to the presence of an 
inner baryonic matter. If the decay pattern we have considered here is 
realistic, 
the lifetime of SDM is close to $50\tau$. Considering the uncertainties in 
determination of fluxes and in the simulation, one can conclude that present 
data is compatible with a mass $m_X \sim 10^{22}eV-10^{24}$ and a lifetime 
${\tau}_X \sim 10\tau - 100\tau$. Our results are also consistent with the 
latest upper limit on the flux of high energy neutrino from Lake Baikal 
experiment~\cite{nubaikal}. Nonetheless complex and mostly unknown physics 
of neutrinos reduces the reliability of high energy neutrino constraints.

\subsection{Constraints from Other Cosmological Data}
Before concluding this section we discuss some of other observable 
consequences of a decaying SDM and constraints they put on the parameter 
space of SDM models. 

A relatively short living SDM can distort CMB. This issue has been studied 
and constrained after recent WMAP observation of CMB 
anisotropy~\cite{wmapdmcons}. Their constraints from WMAP 
spectrum of CMB are based on the modification of equation of state of the 
Universe (see next section) without considering the energy dissipation of 
remnant. The lower limit on the lifetime of SDM at $95.4\% C.L.$ is 
${\tau}_X \gtrsim 52 Gyr \approx 4 \tau$ and at $68\% C.L.$, 
${\tau}_X \gtrsim 123 Gyr \approx 9 \tau$. 
These results even without considering the energy dissipation are 
compatible with the simulation discussed in the previous section and our 
constraints are even somehow more stringent than what obtained 
in ~\cite{wmapdmcons}. 

\begin{figure}[t]
\begin{center}
\psfig{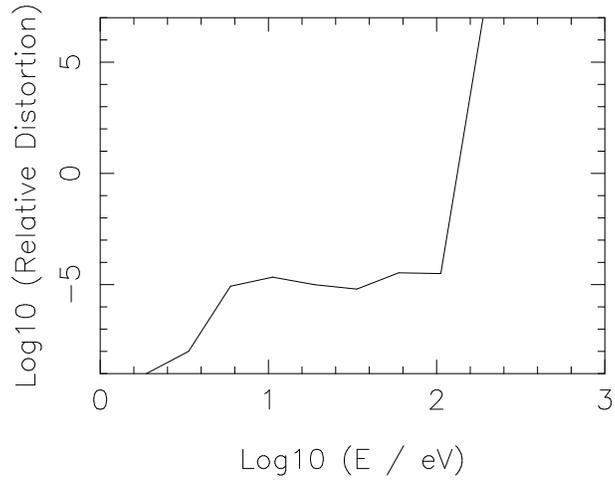}
\caption{Fraction of distortion in photon distribution with respect to a stable DM in energies close to pick of the CMB (Energies are in eV).\label{fig:cmbdistor}}
\end{center}
\end{figure}

In fact we have also calculated the distortion due to SDM decay for a 
homogeneous universe. Fig.\ref{fig:cmbdistor} shows the relative distortion 
of photon background around 
CMB energy for a homogeneous universe dominated by SDM with respect to a 
stable dark matter. The distortion around 
maximum of CMB spectrum is very small, less than  $1$ to $10^8$ parts for 
$E \lesssim 3 eV$. The exact numerical values however depends somehow on the 
cross-section cuts at low energies. Nevertheless, the 
contribution of SDM remnants at low energies close to maximum of the CMB 
spectrum is much smaller than other foreground sources like galaxies and 
galaxy clusters. The study of distortion in CMB anisotropy is more 
complicated. A simple 
estimation can be obtain by multiplying the uniform distortion be an average 
over-density. For an over-density of order $\sim 100$ at the scale of 
clusters, 
the expected distortion at small angle (large $l$) is $\lesssim 10^{-6}$, 
much smaller than resolution of present and near future CMB anisotropy 
experiments.

The other important cosmological constraint is the increase in the entropy 
due to changing of CDM to Hot Dark Matter (HDM). Our simulation shows that 
the increase in the entropy of electrons, protons and photons are completely 
negligible. There is a small increase in $e^+$ and $p^-$ entropy, but much 
smaller than the total entropy of the Universe and compatible with 
observations.

As the mass scale of SDM is expected to be close to GUT scale, it has been 
suggested~\cite{houribbn}~\cite{revorg} that their decay  may be able 
to generate additional baryon and lepton asymmetry. The rate of baryonic (or leptonic) number production by decay of SDM in 
comoving frame can be expressed as: 
\be
\frac {d(n_b - n_{\bar b})}{dt} + 3 \frac {\dot a (t)}{a (t)} (n_b - 
n_{\bar b})= \frac {\varepsilon n_{dm}}{\tau}. \label {nbaryon}
\ee
where $\varepsilon$ is the total baryon number violation per decay. The 
solution of this equation is:
\bea
\Delta (n_b - n_{\bar b}) & = & \varepsilon n_{dm}(t_0) (1 - \exp (- \frac 
{t - t_0}{\tau})) \frac {(1+z_0)^3}{(1+z)^3}. \label {nbs}\\
\Delta B & \equiv & \frac {\Delta (n_b - n_{\bar b})}{2 g_* n_\gamma} = 
\frac {\varepsilon n_{dm}(t_0)}{2 g_* n_\gamma (t_0)} (1 - \exp (- \frac 
{t - t_0}{\tau})) \frac {(1+z)}{(1+z_0)}. \label {db}
\eea
If $t_0 = t_{dec}$, $\frac {n_{dm}(t_0)}{n_\gamma (t_0)} \sim 10^{-22}$ 
(for $m_{dm} = 10^{24} eV$). Therefore $\Delta B \sim 10^{-22} \varepsilon$ 
at $z = 0$. 
As $\varepsilon$ can not be larger than total multiplicity, $\sim 1000$, 
$\Delta B \lesssim 10^{-19}$, i.e. much smaller than primordial value 
$\sim 10^{-10}$. For $\varepsilon = 0.1$ at all energies, 
$\varepsilon_{tot} = 0.1 {\mathcal M_{tot}}$. This leads to a smaller 
$n_{\bar p}$ i.e. larger total baryonic number, but the change is very 
small. The same is true for leptonic number, but energy density of 
leptons with respect to anti-leptons increases by an amount comparable to 
$\varepsilon$.

\section{Equation of State of the Universe}
We continue the odyssey of a superheavy dark matter by considering its 
signature on one of the hottest and most mysterious topics of physics and 
cosmology today: the Equation of State of the Universe (ESU).

Here we show that the decay of dark matter has an effect very similar to 
dark energy. In fact it is easy to see the reason crudely. Assuming that the 
decay remnants stay relativistic, cosmological evolution equation close to 
present time is:
\be
H^2 = \frac {8\pi G}{3} \biggl ({\rho}_{X}(t_0)\frac{a_0^3}{a^3} + 
{\rho}_{X}(t_0)\frac{a_0^4}{a^4}(1 - e^{-\frac {(t-t_0)}{{\tau}_X}}) + 
e^{-\frac {(t-t_0)}{{\tau}_X}} + {\rho}_{hot} + {\rho}_q \biggr ) 
\label {apphubbledec}
\ee
where $\rho$ indicates the density and from now on the subscript $q$ is used 
for quintessence. In this section we assume that quintessence term is a 
cosmological constant. Time $t_0$ is an arbitrary initial time. Equation 
(\ref{apphubbledec}) should be compared with the evolution equation for a 
cosmology with a stable DM:
\be
H^2 = \frac {8\pi G}{3} \biggl ({\rho}_{X}(t_0)\frac{a_0^3}{a^3} + 
{\rho}_{hot} + {\rho}_q \biggr ) \label {apphubble}
\ee
If $t-t_0 \ll {\tau}_X$ and the first cosmology is treated as the second, 
the observer find a slightly smaller density for DM but a growing dark 
energy i.e. $w_q \lesssim -1$ where $w_q$ determines the equation of state 
defined as:
\be
P = w \rho \label {wdef}
\ee
$P$ is pressure. For dark matter $w = 0$, for hot matter $w = 1/3$ and for a 
cosmological constant $w = -1$. A more detailed proof can be found in 
~\cite{snhouri}.

\begin{figure}[t]
\begin{center}
\psfig{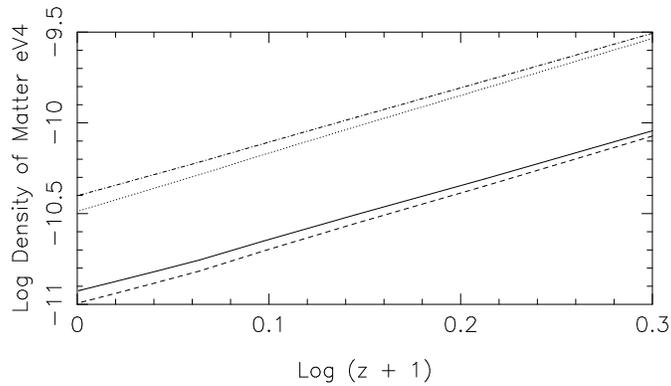}
\caption{Energy density of the Universe. Solid line $\Omega_{\Lambda} = 0.7$ and stable DM; dashed line the same cosmology with ${\tau}_X = 5 \tau$; dash dot line $\Lambda = 0$ and stable DM; dot line $\Lambda = 0$ and ${\tau}_X = 5 \tau$. Dependence on the mass of DM is negligible.\label{fig:t00}}
\end{center}
\end{figure}

Fig.\ref{fig:t00} shows the evolution of $\rho (z)$ the density of CDM+HDM 
at low and medium redshifts in a flat universe with and without a 
cosmological constant and when DM is stable or it is decaying. As expected, 
the effect of SDM decay is more significant in a matter 
dominated universe i.e. when $\Lambda = 0$. For a given cosmology, the 
lifetime of SDM is the only parameter that significantly affects the 
evolution of $\rho$ and the difference between models with 
$M_{X} = 10^{12} eV$ and $M_{X} = 10^{24} eV$ is only $\approx 0.4\%$. 
Consequently, in the following we neglect the effect of DM mass.
We have tried~\cite{snhouri} to see if we can find the finger print of a 
decaying SDM in SN Type-Ia data which is the most direct way to study the 
equation of state of the Universe~\cite{snmeasur}~\cite{newsn}. The 
measurement is based on observation of maximum apparent magnitude of 
SN Type-Ia's light-curve. 
After correction for various observational and intrinsic variations 
like K-correction, width-luminosity relation, metalicity, reddening and 
Galactic 
extinction, it is assumed that their magnitude is universal. The difference 
in apparent magnitudes of SNs is then only related to difference in distance 
and consequently to cosmological parameters. The apparent magnitude of an 
object $m (z)$ is related to its absolute magnitude $M$:
\be
m (z) = M + 25 + 5 \log D_L \label {mag}
\ee
where $D_L$ is the Hubble-constant-free luminosity distance:
\be
D_L = \frac {(z + 1)}{\sqrt{|\Omega_R|}} {\mathcal S} \biggl (\sqrt{|\Omega_R|} 
\int_{0}^{z} \frac {dz'}{E (z')}\biggr ) \label {Dl}\\
\ee
\bea
{\mathcal S}(x) =
\begin {cases}
\sinh (x) & {\Omega}_R > 0, \nonumber \\
x & {\Omega}_R = 0, \nonumber \\
\sin (x) & {\Omega}_R < 0. \label {sdef}
\end {cases}
\eea

\bea
E (z) & = & \frac {H (z)}{H_0}. \label {ez}\\
H^2 (z) & = & \frac {8\pi G}{3} T^{00} (z) + \frac {\Lambda}{3}. \label {hz}
\eea
In (\ref{hz}) we used energy-momentum tensor in place of $\rho$ to 
distinguish between ideal gas approximation and the general case where 
matter components are in interaction and their distribution is not 
necessarily thermal. This is the case when DM decays at late time and 
the distribution of remnants remains non-thermal. 
We restrict fits to flat Cosmologies and fit cosmological models to 
published high redshift SN observations. Cosmological models with and 
without Cosmological Constant and stable or decaying DM are fitted to the 
data. We use minimum-$\chi^2$ method for fitting. Universal absolute 
magnitude $M$ is considered as a free parameter and $\chi^2$ of each 
model is minimized with respect to it. Following aprioris are applied to 
the present density of dark energy:
\be
2.38 \times 10^{-11} \leqslant \rho_\Lambda \equiv \frac {\Lambda}{8\pi G} 
\leqslant 3.17 \times 10^{-11}eV^4 \label {decons}
\ee
We use $\rho_\Lambda$ rather than $\Omega_\Lambda$ because the latter 
depends on the equation of state and lifetime of the dark matter. The 
range of $\rho_\Lambda$ given here is equivalent to 
$0.6\leqslant \Omega_{\Lambda}^{eq} \leqslant 0.8$ for a stable CDM and 
$H_0 = 70$ $km$ $Mpc^{-1} \sec^{-1}$ (This is the value used as initial 
input to the simulation of SDM decay (See ~\cite{houri} for details of 
initial conditions). We use $\Omega_{\Lambda}^{eq}$ 
notation to distinguish between this quantity which is obtained from 
simulation using (\ref{hz}) and the input 
$\Omega_\Lambda = \Lambda / 3{H_0}^2$.

\begin{figure}[t]
\begin{center}
\psfig{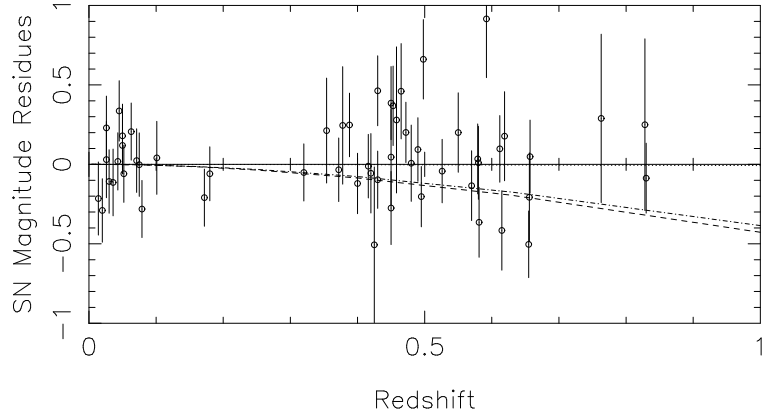}
\caption{Best fit residues with $\Omega_{\Lambda} = 0.7$, ${\tau}_X = 5
\tau$. It leads to $\Omega_{\Lambda}^{eq} = 0.73$. The curves correspond to
residue for stable DM with ${\Omega}_{\Lambda}^{eq} = \Omega_{\Lambda} = 0.7$
(doted); $\Lambda = 0$ and ${\tau} = 5 \tau$ (dashed); $\Lambda = 0$, stable
DM (dash-dot).\label{fig:bestres}}
\end{center}
\end{figure}

Fig.\ref{fig:bestres} shows the residues of the best fit to SDM simulation. 
Although up to $1$-$\sigma$ uncertainty all models with stable or decaying 
DM with $5 \tau \lesssim {\tau}_X \lesssim 50 \tau$ and $0.68\lesssim 
\Omega_{\Lambda}^{eq} \lesssim 0.72$ are compatible with the data, a 
decaying DM with ${\tau}_X \sim 5 \tau$ systematically fits the data 
better than stable DM with the same $\Omega_{\Lambda}$. Models with 
$\Lambda = 0$ are ruled out with more than $99\%$ confidence level.

In fitting the results of DM decay simulation to the data we have directly 
used the equation (\ref{hz}) without defining any analytical form for the 
evolution of $T^{00}(z)$. To be able to compare our results directly with 
other works, we have also fitted an analytical model to the simulation. 
It includes a stable DM and a quintessence matter. Its evolution equation is:
\be
H^2 (z) = \frac {8\pi G}{3} (T^{00}_{st} + {\Omega}_q (z + 1)^{3 (w_q + 1)}). \label {quineq}
\ee
The term $T^{00}_{st}$ is obtained from our simulation when DM is stable. 
In addition to CDM, it includes a small contribution from hot components i.e 
CMB and relic neutrinos. Therefore in this model all the effects of a 
decaying 
DM is encapsulated in the quintessence model. The time/redshift variation of 
dark energy is thus due to decaying DM. For a given $\Omega_{\Lambda}$ and 
$\tau$, the quintessence term is fitted to:
\be
T^{00} - T^{00}_{st} + \frac {\Lambda}{8\pi G}  \label{t00diff}
\ee
Note that the exact equivalent model is:
\be
H^2 (z) = \frac {8\pi G}{3} ((1 - {\Omega}_q) (z + 1)^3 + 
{\Omega}_q (z + 1)^{3 (w_q + 1)}). \label {quinanal}
\ee
However, because (\ref{quinanal}) depends only on one density, the 
minimization of $\chi^2$ of the fit in this model have a trivial solution 
with $w_q = -1$, $\Omega_q = {\Omega}_{\Lambda}$. 
Non-trivial solutions depend on both $w_q$ and $\Omega_q$ which are 
degenerate with infinite number of solutions. The model we have 
used here generates a very good equivalent model to SDM with 
less than $2\%$ error, but because CDM and quintessence terms are not fitted 
together, $\Omega$ is not exactly $1$. 

\begin{table}[t]
\caption{Cosmological parameters from simulation of a decaying DM and 
parameters of the equivalent quintessence model. $H_0$ is in $km$ $Mpc^{-1}
\sec^{-1}$ and correspond to $H_0$ after fitting matter and quintessence 
densities.\label{tab:quineq}}
\vspace{0.2cm}
\begin{center}
\begin {footnotesize}
\begin{tabular}{|c|c|c|c|c|c|c|c|c|c|}
\hline
 &
\multicolumn {3}{c|}{Stable DM} & 
\multicolumn {3}{c|}{${\tau}_X = 50 \tau$} & 
\multicolumn {3}{c|}{${\tau}_X = 5 \tau$} \\
\hline
\raisebox{0pt}[12pt][6pt]{${\Omega}_{\Lambda}$} &
\raisebox{0pt}[13pt][7pt]{$0.68$} &
\raisebox{0pt}[13pt][7pt]{$0.7$} &
\raisebox{0pt}[13pt][7pt]{$0.72$} &
\raisebox{0pt}[13pt][7pt]{$0.68$} &
\raisebox{0pt}[13pt][7pt]{$0.7$} &
\raisebox{0pt}[13pt][7pt]{$0.72$} &
\raisebox{0pt}[13pt][7pt]{$0.68$} &
\raisebox{0pt}[13pt][7pt]{$0.7$} &
\raisebox{0pt}[13pt][7pt]{$0.72$}\\
\hline
\raisebox{0pt}[12pt][6pt]{$H_0$}  
 & \raisebox{0pt}[12pt][6pt]{$69.953$}
 & \raisebox{0pt}[12pt][6pt]{$69.951$} & \raisebox{0pt}[12pt][6pt]{$69.949$}
 & \raisebox{0pt}[12pt][6pt]{$69.779$} & \raisebox{0pt}[12pt][6pt]{$69.789$}
 & \raisebox{0pt}[12pt][6pt]{$69.801$} & \raisebox{0pt}[12pt][6pt]{$68.301$}
 & \raisebox{0pt}[12pt][6pt]{$68.415$} & \raisebox{0pt}[12pt][6pt]{$68.550$}\\
\hline
\raisebox{0pt}[12pt][6pt]{$\Omega_{\Lambda}^{eq}$}
 & \raisebox{0pt}[12pt][6pt]{$0.681$}
 & \raisebox{0pt}[12pt][6pt]{$0.701$} & \raisebox{0pt}[12pt][6pt]{$0.721$}
 & \raisebox{0pt}[12pt][6pt]{$0.684$} & \raisebox{0pt}[12pt][6pt]{$0.704$}
 & \raisebox{0pt}[12pt][6pt]{$0.724$} & \raisebox{0pt}[12pt][6pt]{$0.714$}
 & \raisebox{0pt}[12pt][6pt]{$0.733$} & \raisebox{0pt}[12pt][6pt]{$0.751$}\\
\hline
\raisebox{0pt}[12pt][6pt]{$\Omega_q$}
 & \raisebox{0pt}[12pt][6pt]{-}
 & \raisebox{0pt}[12pt][6pt]{-} & \raisebox{0pt}[12pt][6pt]{-}
 & \raisebox{0pt}[12pt][6pt]{$0.679$} & \raisebox{0pt}[12pt][6pt]{$0.700$}
 & \raisebox{0pt}[12pt][6pt]{$0.720$} & \raisebox{0pt}[12pt][6pt]{$0.667$}
 & \raisebox{0pt}[12pt][6pt]{$0.689$} & \raisebox{0pt}[12pt][6pt]{$0.711$}\\
\hline
\raisebox{0pt}[12pt][6pt]{$w_q$}
 & \raisebox{0pt}[12pt][6pt]{-}
 & \raisebox{0pt}[12pt][6pt]{-} & \raisebox{0pt}[12pt][6pt]{-}
 & \raisebox{0pt}[12pt][6pt]{$-1.0066$} & \raisebox{0pt}[12pt][6pt]{$-1.0060$}
 & \raisebox{0pt}[12pt][6pt]{$-1.0055$} & \raisebox{0pt}[12pt][6pt]{$-1.0732$}
 & \raisebox{0pt}[12pt][6pt]{$-1.0658$} & \raisebox{0pt}[12pt][6pt]{$-1.0590$}\\
\hline
\raisebox{0pt}[12pt][6pt]{$\chi^2$}
 & \raisebox{0pt}[12pt][6pt]{$62.36$}
 & \raisebox{0pt}[12pt][6pt]{$62.23$} & \raisebox{0pt}[12pt][6pt]{$62.21$}
 & \raisebox{0pt}[12pt][6pt]{$62.34$} & \raisebox{0pt}[12pt][6pt]{$62.22$}
 & \raisebox{0pt}[12pt][6pt]{$62.21$} & \raisebox{0pt}[12pt][6pt]{$62.22$}
 & \raisebox{0pt}[12pt][6pt]{$62.15$} & \raisebox{0pt}[12pt][6pt]{$62.20$}\\
\hline
\end{tabular}
\end{footnotesize}
\end{center}
\end{table}
Parameters of models in the $1$-$\sigma$ distance of the best fit are 
summarized in Table \ref{tab:quineq}. The results for ${\tau}_X = 5 \tau$ 
models are surprisingly close to the results obtained recently by fitting 
the best SN light-curves observed by HST~\cite{hstsn}. Unfortunately the 
errors of both fits are too large to make any definitive conclusion. 
Nevertheless, there is very small chance 
that closeness of mean values be just accidental. The lifetime for the 
best-fit models is somehow smaller than the lower limit we found in 
Sec.\ref{sec:evol}. 
However, one should not forget that hadronization we have considered is 
maximal. i.e. we have considered that all the remnants of the decay of SDM 
is visible and has the same baryonic fraction as low energy hadronization. 
If part of the remnants consist of a lighter dark matter e.g. LSP 
(neutralino) the lower limit on the lifetime decreases.

Conclusion we can make from this section is that there is probably a finger 
print of a decaying 
dark matter in the present data. Evidently this is not the only model which 
can explain a $w_q \lesssim -1$. But most other models need a fine-tuning.
They have either unconventional kinetic terms~\cite{negqw} or a negative 
potential which in the context of SUSY models (before breaking) can not be 
obtained, or unconventional equation of state like a 
Chaplygin Gas~\cite{chapgas}. Other scalar field models with multiple-field 
contents or what is called a phantom matter which has a negative kinetic 
energy have been also suggested~\cite{quinmult}~\cite{altpoten}.

\section{Quintessence} \label{sec:quin}
Up to now we used cosmologies with a dark energy (or Cosmological Constant) 
without talking about the nature of this mysterious term in Einstein 
equation (For review see ~\cite{quinrev} and references therein).

Cosmological Constant has been added by Einstein to his equation to be able 
to have a static solution (For a historical review see ~\cite{quinhist}). 
Later however, it was proved by Friedmann that this solution is 
unstable~\cite{friedman}. 
George Gamov has written that he once heard from Einstein to call 
Cosmological Constant his {\it greatest blunder}. But in a letter to 
Einstein, Lema\^ire says that it is a genius idea and interprets it as being 
the {\it Vacuum Energy}. This name is the origin of a significant confusion 
and many speculations and doubts even today. We come back to this point 
later. 
In their famous book {\it Gravitation}, Misner, Thorne and Wheeler call it 
{\it Pandora Box} and consider it only exceptionally.

Today we know that Cosmological Constant or dark energy is the dominant 
contents of the Universe. According to SN data: ${\Omega}_{Lambda} = 
0.75^{+0.07}_{-0.06}$~\cite{hstsn}, and from CMB anisotropy measurement: 
${\Omega}_{Lambda} \approx 0.73$. In a universe very close to flat this 
means that dark energy contribution is more than $70\%$ of the total energy 
contents of the Universe. However, as the density of DE had barely changed 
presumably since after inflation, this means that at that time its value was 
$\sim 10^{45}$ to $\sim 10^{100}$ times (depending on inflation scale) 
smaller than matter density. Such a small value became nonetheless dominant 
after galaxy formation, a good luck for us, otherwise perturbations 
couldn't grow to make structures we see in present Universe including 
ourselves. The situation is worth if the origin of Cosmological Constant 
is related to quantum gravity. For instance if it is the vacuum 
expectation value of a quantity at quantum gravity scale - as 
Lema\^itre suggested - its natural value should be 
$\sim {M_p}^4 \approx 10^{112} eV^4$, i.e. $\sim 10^{123}$ times larger than 
observed value. The unexpectedly small value of Cosmological Constant 
- if Lema\^itre interpretation is correct - 
is called first cosmological constant problem. The fine-tuning such that 
it become dominant only after galaxy formation is called second cosmological 
constant or coincidence problem (see ~\cite{quinrev} and references therein).

The first problem can be solved or soften if dark energy is not vacuum energy 
but comes from another form of matter. This idea is the basis for most of 
suggested models. A reasonable solution for the second problem logically 
seems to be a direct relation between dark matter and dark energy such that 
somehow they control each other and what we consider to be a 
{\it coincidence} is something inherent to the nature of these entities.

A number of dark energy models have been made based on this line of thinking.
In ~\cite{quinx} a matter component with smooth equation of state i.e 
$w_q \sim -0.3$ and dependence on the total energy has been proposed. Another 
possibility is an interaction between dark matter and dark energy. 
Various type of interactions have been investigated. one of them is an 
asymptotic scaling law between density of DE and DM. In this  
model due to a dissipative interaction between dark matter and quintessence 
scalar field $\phiq$, the relative density of dark matter and dark energy
${\rho}_{CDM} / {\rho}_q$ approaches a constant 
value~\cite{interac0}~\cite {interac1}~\cite {interac2}. 
A class of potentials $V_q ({\phi}_q)$ have been found such that the 
equation of state have a solution satisfying this {\it strong 
coincidence scaling}~\cite {interac0}. Constraints on this model from 
nucleosynthesis, leads to $w_q \gtrsim -0.7$ which is only marginally 
compatible with WMAP data and far from publicly available SN-Ia data which 
prefers $w_q \sim -1$. 

Interaction between DM and DE have been extensively studied in the context 
of traditional quintessence models with tracking solutions and 
$w_q > -1$~\cite {interac3}. It has been shown that these models have 
lagrangians equivalent to Brans-Dicke lagrangian with power law potential 
and consequently behave like a {\it Fifth Force}. 
Modification of the CMB anisotropy spectra by such interactions is 
observable and put stringent constraints on their parameters. 

Models with a time dependent DM mass due to interaction between 
quintessence scalar and dark matter have been also 
considered~\cite {massvar}. Coupling between two fields in this 
class of models increases the parameter space for both and reduces by 
orders of magnitudes the amount of fine tuning. However, there are strong 
limits on the variation of fundamental parameters including DM mass. 
Moreover, in these models the largest amount of variation happens around 
and after matter domination epoch. Consequentlt the mass variation must 
leave an imprint on the CMB and large structure formation which has not 
been observed.

Most quintessence models suffer also from difficulties regarding particle
physics model for the quintessence scalar field with proper 
mass ~\cite {quinpot}. Quintessence field is usually considered to be an 
axion-like particle with high-order, non-renormalizable interactions 
with SM (or its super-symmetric extension) fields. However, any 
supergravity induced interaction between 
${\phi}_q$ and other scalars with VEV of the order of Planck mass can 
increase the very tiny mass of the ${\phi}_q$ - in most models 
$m_q \sim H_0 \sim 10^{-33} eV$ - unless a discrete global symmetry 
prevents their contribution to the mass~\cite {massrenor}(see also 
\cite{axionmod} for some solutions for this problem). In summary no ideal 
solution for coincidence problem has yet been found.

In this section we describe a model for the dark energy somehow different 
from previous quintessence models~\cite{houriquin}. We assume that DE is 
the result of condensation of a scalar field produced during very slow 
decay of the superheavy dark matter which we have studied in the previous 
sections. In traditional quintessence models the 
scalar field is produced during inflation or reheating period in large 
amount such that for controlling its contribution to the total energy of the 
Universe today, its potential must be decreasing since that time. Usually 
the potential is a negative exponential, sum of two exponentials, or a 
negative power polynimial function and their parameters must be somehow 
fine-tuned~\cite {quinpot}. In the 
present model very small production rate of the scalar field due to very 
small decay rate of SDM replaces the fine-tuning of the potential and 
practically any scalar field even without a self-interaction has a tracking 
solution for a large part of its parameter space. We will see that soon 
after production of SDM, $\phiq$ behaves like a cosmological constant 
without need for fine-tuning of parameters. As we have seen in the 
previous section, subsequently the decay of SDM imitates a universe with a 
quintessence field 
for which $w_q$ is slightly smaller than $-1$. This is exactly what has been 
observed ! Another advantage of this model is that there would not be any 
future horizon because one day SDM will completely decay and automatically 
changes the equation of state of the quintessence field. The existence of 
a future horizon in an accelerating universe is problematic because some 
particle physics models specially string theory lack a well defined vacuum 
in a de Sitter space with future horizon. We also show that large mass and 
lifetime of SDM is crucial for making this model a proper quintessence model. 

What we present here is based on the assumption that late time decoherence 
of quintessence field is possible. It has been shown in case of inflaton 
that decoherence is only possible for modes with a wavelength larger than 
horizon. This put an upper limit on the mass of the quintessence, later the 
decoherence time, smaller the mass upper limit. If SDM is produced 
during preheating ~\cite {grapro} just after the end of the inflation 
presumably at scales $\sim 10^{14}eV-10^{16}eV$ which correspond to:
\be
H \sim 10^{-6}eV-10^{-4}eV    \label {usize}
\ee
the permitted mass range is $m_q \lesssim 
10^{-6}eV$~\cite{decohereinf0}~\cite{decohereinf1}. When the 
size of the Universe get larger, $\phiq$ stops decohering. This also helps 
having a very small dark energy density. If the preheating/reheating had 
happened when the Hubble Constant was smaller, then $m_q$ also must be 
smaller to have long wavelength modes which can decohere. The issue of 
decoherence is very complex and needs more investigation.

\subsection {Co-Evolution of Decaying SDM and Quintessence Field} \label {sec.qmodel}
Consider that just after inflation among the field contents of the 
Universe there is $\phix$ a superheavy, meta-stable dark matter (SDM) which 
decouples from the rest of the {\it primordial soup} since very early time. 
These are the same properties we assumed for $X$ particles in the previous 
sections. We don't consider 
other fields in detail. The only constraint on the other fields is that 
they must consist of light species including baryons, neutrinos, photons, 
and light dark matter - by light we mean with respect to X. For simplicity 
we assume that $X$ is a scalar field $\phix$. If $\phix$ is a 
spinor or vector the general conclusions presented here does not change. 
A very small part of $\phix$ decay remnants is considered to be a scalar 
field $\phiq$ with negligibly weak interaction with other fields.

The effective lagrangian can be written as:
\be
{\mathcal L} = \int d^4 x \sqrt{-g} \biggl [\frac {1}{2} g^{\mu\nu} 
{\partial}_{\mu} \phix {\partial}_{\nu} \phix + \frac {1}{2} g^{\mu\nu} 
{\partial}_{\mu} \phiq {\partial}_{\nu} \phiq - V (\phix, \phiq, J) 
\biggr ] + {\mathcal L}_J \label {lagrange}
\ee
The field $J$ presents collectively other fields. 
The term $V (\phix, \phiq, J)$ 
includes all interactions including self-interaction potential for 
$\phix$ and $\phiq$:
\be
V (\phix, \phiq, J) = V_q (\phiq) + V_x (\phix) + g {\phix}^m {\phiq}^n + 
W (\phix, \phiq, J) \label {potv}
\ee
The term $g {\phix}^m {\phiq}^n$ is important because it is responsible for 
annihilation of $X$ and back reaction of quintessence field to SDM. 
$W (\phix, \phiq, J)$ presents interactions which contribute to the decay 
of $X$ to light fields and to $\phiq$ (in addition to what is shown 
explicitly in (\ref{potv})). The very long lifetime of $X$ constrains 
this term and $g$. They must be strongly suppressed. For $n = 2$ and 
$m = 2$ the term proportional to $g$ contributes to the mass of $\phix$ and 
$\phiq$. Because of the huge mass of $\phix$ (which 
must come from another coupling) and its very small occupation number, 
we can use classical limit i.e. 
$<{\phix}^2> \sim 2 {\rho}_x / {m_x}^2$. For sufficiently small $g$, the 
effect of this term on the mass of the SDM is very small. 
We discuss the r\^ole of this term in detail later. If the interaction of 
other fields with $\phiq$ is only through the exchange of $X$ (for instance 
due to a conserved symmetry shared by both), the huge mass of $X$ 
suppresses the interaction and therefore the modification of $m_q$. 
If $X$ is a spinor, the lowest order (Yukawa) interaction term in 
(\ref {lagrange}) is $g \phiq \bar {\psi} \psi$. In the classical treatment 
of $X$:
\be
\bar {\psi} \psi \sim \frac {{\rho}_x}{m_x} \label {spinor}
\ee
The lagrangian (\ref {lagrange}) leads to following system of equations for 
the fields:
\bea
\dot{\phiq} [\ddot{\phiq} + 3H \dot{\phiq} + {m_q}^2 \phiq + 
\lambda {\phiq}^3] & = & -2g \dot{\phiq}\phiq 
\biggl (\frac {2 {\rho}_x}{{m_x}^2}\biggr ) + {\Gamma}_q{\rho}_x 
\label {phiqe} \\
\dot {{\rho}_x} + 3H {\rho}_x & = & - ({\Gamma}_q + {\Gamma}_J)
{{\rho}_x} - {\pi}^4 g^2 \biggl (\frac {{{\rho}_x}^2}{{m_x}^3} - 
\frac {{{\rho}_q}'^2}{{m_q}^3}\biggr ) \label {xeq} \\
\dot {{\rho}_J} + 3H ({\rho}_J + P_J) & = & {\Gamma}_J {{\rho}_x} 
\label {jeq} \\
H^2 & \equiv & \biggl (\frac {\dot{a}}{a}\biggr )^2 = \frac {8\pi G}{3} 
({\rho}_x + {\rho}_J + {\rho}_q) \label {heq} \\
{\rho}_q & = & \frac {1}{2} {m_q}^2 {\dot{\phiq}}^2 + 
\frac {1}{2} {m_q}^2 {\phiq}^2 + \frac {\lambda}{4} {\phiq}^4 
\label {phidens}
\eea
Eq. (\ref {xeq}) is Boltzmann equation for $X$ particles and its right 
hand sides can be obtained from detail calculation of its annihilation and 
reproduction due to term $g {\phix}^2 {\phiq}^2$ in the 
lagrangian~\cite{houriquin}. ${{\rho}_q}'$ is the 
density of quintessence particles (not the classical field $\phiq$) 
with an average energy larger than $m_x$ in the local inertial frame. Only 
interaction between these 
particles contribute to the reproduction of SDM. ${\Gamma}_q$ and 
${\Gamma}_J$ are respectively the decay width of $X$ to $\phiq$ and to other 
species. The effect of decay lagrangian $W (\phix, \phiq, J)$ appears as 
$({\Gamma}_q + {\Gamma}_J){\rho}_x$ which is the decay rate of $X$ 
particles.

At very high temperatures when ${\rho}_x \gg {\pi}^4g^2{m_x}^3 \Gamma$, the 
annihilation and reproduction terms in (\ref {xeq}) are dominant. $X$ 
particles however are non-relativistic up to temperatures close to their 
rest mass. Quintessence scalar particles at this time are relativistic and 
therefore their density falls faster than SDM density by a factor of 
$a (t)$. 
The probability of self-annihilation also decreases very rapidly. 
Consequently, 
from very early time only the decay term in (\ref {xeq}) is important. The 
dominance of annihilation/reproduction can happen only if the production 
temperature of $X$ particles i.e. preheating/reheating temperature is very 
high. Such scenarios however can make dangerous amount of gravitinos ~\cite 
{gravprod}. A lower reheat temperature does not however compromise the 
production of SDM because it has been shown ~\cite {wimprod} that 
even with a very low reheating temperature they can be produced. 
Another reason for this simplification is that we are interested in the 
decohered modes of $\phiq$. Self-annihilation of $X$ particles makes 
$\phiq$ particles which are highly relativistic and don't participate in the 
condensate modes.

Equations (\ref {phiqe}) to (\ref {phidens}) are highly non-linear and 
coupled. There are however two asymptotic 
regimes which permit an approximate analytical treatment. The first one 
happens very early just after production of $X$ particles presumably 
after preheating ~\cite {wimpzilla}~\cite {grapro} and decoherence of 
$\phiq$'s long wavelength modes. In this epoch $\phiq \sim 0$ and can be 
neglected. The other regime is when comoving time variation of $\phiq$ is 
very slow and one can neglect $\ddot {\phiq}$. We show that the first 
regime leads to a saturation (tracking) solution where 
$\phiq \rightarrow cte.$ It then can be treated as the initial condition 
for the second regime when $\phiq$ changes slowly.

Neglecting the last term in the right hand side of (\ref {xeq}), this 
equation has a straightforward solution:
\be
{\rho}_x (t) = {\rho}_x (t_0) e^{-\Gamma (t-t_0)} \biggl (\frac {a (t_0)}
{a (t)}\biggr )^3 \label {xsol}
\ee
where $\Gamma \equiv {\Gamma}_q + {\Gamma}_J$ is the total decay width of 
$X$. Initial time $t_0$ is considered to be after production and 
decoupling of $X$. After inserting the solution (\ref {xsol}) and 
neglecting all the terms proportional to $\phiq$, equation (\ref {phiqe}) 
can be solved:
\be
\frac {1}{2}{\dot{\phiq}}^2 (t) \equiv K_q(t) = \biggl (\frac {a (t_0)}
{a (t)}\biggr )^6 \biggl [K_q(t_0) + {\Gamma}_q {\rho}_x (t_0) 
\int_{t_0}^t dt \frac {a^3}{a (t_0)}e^{-\Gamma (t-t_0)} \biggr ]
\label {phiappsol}
\ee
For $a \propto t^k$ the integral term in (\ref {phiappsol}) decreases with 
time (i.e. $\ddot{\phiq} < 0$). This means that after a relatively short 
time $\phiq$ is saturated and its density does not change, in other words 
it behaves like a cosmological constant. Due to quantum effects the initial 
value of $K_q(t_0)$ is positive. Its exact value can only 
be determined by investigating the process of decoherence of $\phiq$. 
Because of $a^{-6}(t)$ factor in (\ref {phiappsol}) however, with 
the expansion of the Universe the effect of this term on $\dot{\phiq}$ 
decreases very rapidly.

During the second regime when $\phiq$ changes very slowly, we can 
neglect $\ddot{\phiq}$ and higher orders of $\dot{\phiq}$. Equation 
(\ref{phiqe}) thus simplifies to:
\be
\dot{\phiq}({m_q}^2 \phiq + \lambda {\phiq}^3) = -2g 
\dot{\phiq}{\phiq} \biggl (\frac {2 {\rho}_x}{{m_x}^2}\biggr ) + 
{\Gamma}_q{\rho}_x \label {phisloworg}
\ee
We expect that self-interaction of $\phiq$ be much stronger than its 
coupling to $X$. Neglecting the first term in the right hand side of 
(\ref{phisloworg}), its $\phiq$-dependent part can be integrated:
\be
\frac {d}{dt} \biggl (\frac {1}{2} {m_q}^2 {\phiq}^2 + \frac {\lambda}{4}
{\phiq}^4 \biggr ) = \frac {dV}{dt}(\phiq) = {\Gamma}_q {\rho}_x 
\label {phislowphi}
\ee
and solved:
\be
V_q (\phiq) = V_q (\phiq (t'_0)) + {\Gamma}_q {\rho}_x (t'_0) 
\int_{t'_0}^t dt \biggl (\frac {a (t'_0)}{a (t)}\biggr )^3 
e^{-\Gamma (t-t'_0)} 
\label {phislowsol}
\ee
Here $V_q$ is the potential energy of $\phiq$. From (\ref {phislowphi}) and 
(\ref {phislowsol}) it is clear that the final value of the potential and 
therefore $\phiq$ energy density is driven by the decay term and not by 
self-interaction. Therefore the only vital condition for this model is the 
existence of a long life SDM and not the potential of $\phiq$. This is very 
different from most quintessence models. In ~\cite{quinhybrid} also a 
${\phi}^4$ potential has been used in the context of hybrid scalar models. 
In this model the dark matter is also a condensed scalar.

In (\ref {phislowsol}) the initial values $t'_0$ and $\phiq (t'_0)$ 
correspond respectively to time and to $\phiq$ in the first regime when 
it approaches to saturation. Similar to 
(\ref {phiappsol}), the time 
dependence of $\phiq$ in (\ref {phislowsol}) vanishes exponentially and the 
behavior of $\phiq$ approaches to a cosmological constant. 

Assuming $a (t) \propto t^k$ and $t_s - t'_0 \ll 1/\Gamma$ where $t_s$ is 
the saturation time, we find:
\be
V (\phiq) - V (\phiq (t'_0)) \sim \frac {{\Gamma}_q{\rho}_x (t'_0)}
{(3k -1)}\biggl (1- \biggl (\frac {t'_0}{t} \biggr)^{(3k -1)} \biggr). 
\label {vtime}
\ee
Defining saturation time as the time when $V (\phiq) - 
V (\phiq (t'_0))$ has 90\% of its final value, if $t_s \ll t_{eq}$ with 
$t_{eq}$ the matter-radiation equilibrium time, $k = 1/2$ and:
\be
t_s \sim 100 t'_0   \label {tsrad}
\ee
For $t_s \gg t_{eq}$, $k = 2/3$ and:
\be
t_s \sim 10 t'_0    \label {tsdm}
\ee

\subsection {Numerical Solution} \label {sec.simul}
A better understanding of the behavior and the parameter space 
of this model needs numerical solution of equations (\ref{phiqe}) to 
(\ref{phidens}). We have also added 
the interaction between various species of the Standard Model (SM) particles 
as explained in Sec.\ref{sec:dissip} to the simulation to be closer to the 
real cosmological evolution and to obtain the equation of state of the 
remnants. We try a number of combination of parameters to find how sensitive 
is the behavior of the quintessence field $\phiq$. 
Parameter space is however degenerate and two models lead to very similar 
results for the quintessence field if:
\be
\frac {f_q}{f'_q} = \frac {z' {\Gamma}' m_x}{z {\Gamma} m'_x} \label {rescal}
\ee
This helps to extend the conclusion to the part of the parameter space which 
is not accessible due to limitations of the numerical simulation.
For the lifetime of $X$ we use ${\tau}_X = 5 {\tau} - 50 \tau$, similar to 
Sec.\ref{sec:dissip}. Results presented here belong to ${\tau}_X = 5$. Our 
test shows that 
increasing ${\tau}_X$ to $50 \tau$ does not significantly modifies the 
main characteristics of dark energy (for more detail see ~\cite{houriquin}).

\begin{figure}[h]
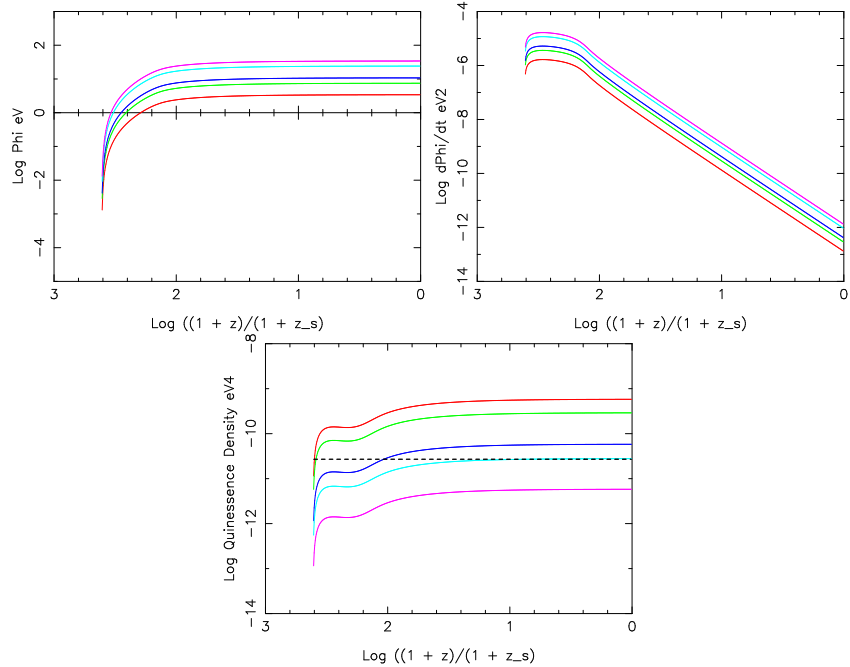

\begin{center}
\psfig{figure=quinphi.eps,angle=-90,width=5.5cm}
\psfig{figure=quindphi.eps,angle=-90,width=5.5cm}
\psfig{figure=quindens.eps,angle=-90,width=5.5cm}
\caption{Evolution of quintessence field (left), its derivative (center) 
and its total energy density (right) for ${\Gamma}_0 \equiv 
{\Gamma}_q/\Gamma = 10^{-16}$ 
(magenta) (see text for details), $5 {\Gamma}_0$ (cyan), $10 {\Gamma}_0$ 
(blue), $50 {\Gamma}_0$ (green), $100 {\Gamma}_0$ (red). Dash line is the 
observed value of the dark energy. $m_q = 10^{-6} eV$, 
$\lambda = 10^{-20}$.
\label {fig:quinevol}}
\end{center}
\end{figure}
Fig.\ref{fig:quinevol} shows the evolution of $\phiq$, its time derivative 
and its total energy density from the end of $X$ production to saturation 
redshift $z_s$. Here we have used as $z_s$ the redshift after which up to 
simulation precision the total energy density of $\phiq$ does not change 
anymore. The result is consistent with the approximate solutions discussed 
earlier. The final density 
of $\phiq$ is practically proportional to ${\Gamma}_q/\Gamma$ which 
encompasses 3 important parameters of the model: The 
fraction of energy of the remnants which changes to $\phiq$, the fraction 
of energy in the long wavelength modes which can decohere and the 
coupling of these modes to the environment which contributes to $\phiq$ 
yield and to the 
formation redshift of the classical quintessence field $\phiq$. 
Therefore the effective volume of the parameter space presented by this 
simulation is much larger and the fine-tuning of parameters are much less 
than what is expected from just one parameter.

\begin{figure}[h]
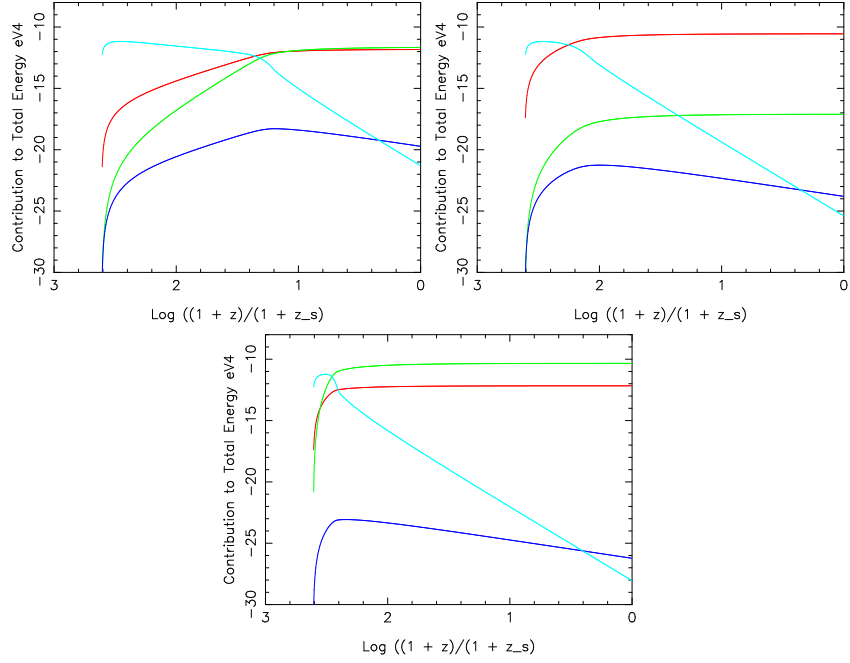

\begin{center}
\psfig{figure=quincontribmass-8.eps,angle=-90,width=5.5cm}
\psfig{figure=quincontrib.eps,angle=-90,width=5.5cm}
\psfig{figure=quincontriblambda-10.eps,angle=-90,width=5.5cm}
\caption{Evolution of the contribution to the total energy density of $\phiq$ 
for ${\Gamma}_0 \equiv {\Gamma}_q/\Gamma = 10^{-16}$ and : Left, 
$m_q = 10^{-8} eV$ and $\lambda = 10^{-20}$; Center, $m_q = 10^{-6} eV$ 
and $\lambda = 10^{-20}$; Right, $m_q = 10^{-6} eV$ and $\lambda = 
10^{-10}$.Curves are: mass (red), self-interaction (green), 
kinetic energy (cyan) and interaction with SDM (blue).
\label {fig:quincontrib}}
\end{center}
\end{figure}
\begin{figure}[h]
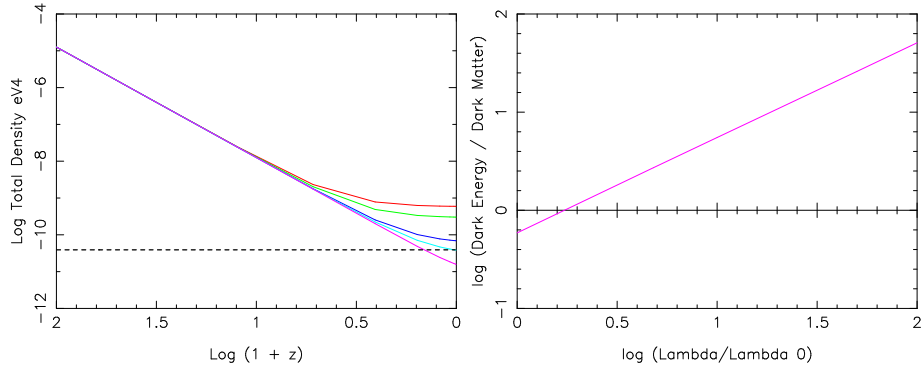

\begin{center}
\psfig{figure=quintotdens.eps,angle=-90,width=6cm}
\psfig{figure=quindmrel.eps,angle=-90,width=6cm}
\caption{Left: Evolution of total density with redshift for 
${\Gamma}_0 \equiv 
{\Gamma}_q/\Gamma = 10^{-16}$ (magenta) (see text for details), 
$5 {\Gamma}_0$ (cyan), $10 {\Gamma}_0$ (blue), $50 {\Gamma}_0$ 
(green), $100 {\Gamma}_0$ (red). Dash line is the observed value of the 
dark energy. $m_q = 10^{-6} eV$, $\lambda = 10^{-20}$. Right: 
Relative density of dark energy and CDM as a function of 
${\Gamma}_q/\Gamma$. The x-axis is normalized to ${\Gamma}_0 \equiv 
{\Gamma}_q/\Gamma = 10^{-16}$.
\label {fig:quintotdens}}
\end{center}
\end{figure}
\begin{figure}[h]
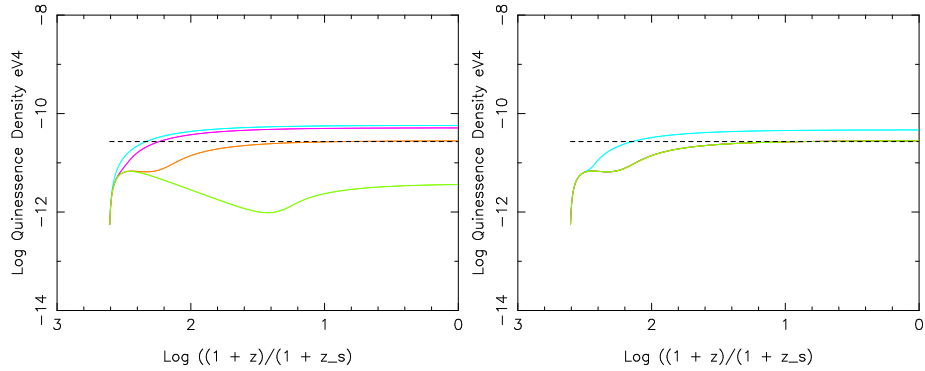

\begin{center}
\psfig{figure=quindenscompmass.eps,angle=-90,width=6cm}
\psfig{figure=quindenscomplambda.eps,angle=-90,width=6cm}
\caption{Quintessence energy density for: Left, 
$m_q = 10^{-3} eV$ (cyan), $m_q = 10^{-5} eV$ (magenta), $m_q = 10^{-6} eV$ 
(red) and $m_q = 10^{-8} eV$ (green), $\lambda = 10^{-20}$; Right, 
$\lambda = 10^{-10}$ (cyan), $\lambda = 10^{-15}$, $\lambda = 10^{-20}$ 
and $\lambda = 10^{-25}$ (green), $m_q = 10^{-6} eV$. The difference between 
quintessence density for the last 3 values of $\lambda$ is smaller than 
the resolution of 
the plot. Dash line is the observed energy density of the dark energy.
\label {fig:quinmass}}
\end{center}
\end{figure}

Fig.\ref{fig:quincontrib} shows the evolution in the contribution of 
different 
terms of the lagrangian (\ref{lagrange}) to the total energy of $\phiq$. 
Very soon after beginning of quintessence field production the potential 
takes over the kinetic energy and the latter begins to decrease. The 
relative contribution of each term and their time of dominance, as this 
figure demonstrates, depends on the model parameters specially on $m_q$ and 
$\lambda$. This plot shows also that changing parameters by orders of 
magnitude does not change the general behavior of 
the model significantly and for a large part of the parameter space the 
final density of quintessence energy is close to the observed value. This 
can also be seen in Fig.\ref{fig:quintotdens} and Fig.\ref{fig:quinmass} 
where the evolution of quintessence energy is shown for various combination 
of parameters. 

\subsection {Perturbations} \label {sec.perturb}
Observations show that the dark energy is smooth and uncorrelated from 
the clumpy dark matter~\cite {sdss}. If its origin is the decay of 
the dark matter, the question arises whether it clumps around dark matter 
halos or has a large scale perturbation which 
is not observed in the present data. We show here that due to special 
characteristics of SDM, $\phiq$ perturbations are very small. 

We use the synchronous gauge metric:
\be
ds^2 = dt^2 - a^2 (t) ({\delta}_{ij} - h_{ij}) dx^i dx^j \label {metric}
\ee
For small spatial fluctuations $\phiq (x,t) = \bar{\phiq}(t) + 
\delta \phiq (x,t)$ where from now on barred quantities are the homogeneous 
component of the field depending only on $t$. We define the same 
decomposition for other fields.

We consider only scalar metric fluctuations $h \equiv {\delta}^{ij} h_{ij}$ 
and neglect other components. Evolution equation for $h$ is:
\be
\frac {1}{2} \ddot {h} + \frac {\dot {a}}{a} \dot {h} = 4 \pi G (4 
\dot {\bar{\phiq}}\dot {\delta \phiq} - 2 \delta V (\phiq, {\rho}_x) + 
\delta {\rho}_x + \delta {\rho}_J + 3 \delta P_J) \label {deltheq}
\ee
where $\delta {\rho}_x$ is the fluctuation of $X$ particles density, 
$\delta {\rho}_J$ and $\delta P_J$ are respectively the collective density 
and pressure fluctuation of other fields. From the lagrangian 
(\ref{lagrange}), the dynamic equation of $\phiq$ is:
\be
{\partial}_{\mu} (\sqrt {-g} g^{\mu\nu} {\partial}_{\nu}\phiq) + 
\sqrt {-g} V'(\phiq,\phix,J) = 0 \label {qgeq}
\ee
This equation and the energy momentum conservation determine the evolution 
of $\delta \phiq (x,t)$:
\bea
& & \dot {\bar {\phiq}}\biggl [ \ddot {\delta \phiq} + {\partial}_i
{\partial}^i (\delta \phiq) + V_q''(\bar {\phiq})\delta \phiq + 
2g \biggl (\frac {2\bar {\rho}_x}{{m_x}^2}\biggr ) \delta \phiq + 
3 \frac {\dot {a}}{a} \dot {\delta \phiq} \biggr ] + \nonumber \\
 & & \hspace {0.5cm}\frac{2g \bar {\phiq}}{{m_x}^2} \biggl [2 \frac {\dot 
{\rho}_x}{\bar {\rho}_x} \delta \phiq + \bar {\phiq} \frac {\dot {\delta 
{\rho}_x}}{\bar {\rho}_x}\biggr ] - \frac {\dot {a}}{a} 
\biggl [h \biggl (\frac {1}{2}\dot {\bar{\phiq}}^2 - V (\bar {\phiq})
\biggr ) - \nonumber \\
 & & \hspace {1cm} 6 \biggl ({V_q}' \delta \phiq + \frac{2g \bar {\phiq} 
\bar {\rho}_x}{{m_x}^2} (2 \delta \phiq + \bar {\phiq} \frac {\delta {\rho}_x}
{\bar {\rho}_x} )\biggr )\biggr ] - 
\frac {\dot {h}}{2}\dot {\bar{\phiq}}^2 = {\Gamma}_q (\delta {\rho}_x - 
\frac {\dot {\delta \phiq}}{\dot {\bar{\phiq}}} \bar {\rho}_x) \nonumber \\
& &
\label {phidoteq}
\eea
Assuming SDM behaves like a pressure-less fluid the energy-momentum tensor 
becomes:
\be
{T_x}^{00} = \bar {\rho}_x + \delta {\rho}_x \quad \quad {T_x}^{0i} = 
\bar {\rho}_x \delta {u_x}^i \quad \quad {T_x}^{ij} = \mathcal {O} 
({\delta}^2) \thickapprox 0 \label {tx}
\ee
where $\delta {u_x}^i$ is the velocity of SDM fluctuations with respect to 
homogeneous Hubble flow. Interaction terms are explicitly included 
in the energy-momentum conservation equation:
\be
{\partial}_0 \biggl (\frac {\delta {\rho}_x}{\bar {\rho}_x} \biggr ) + 
{\partial}_i (\delta {u_x}^i) - \frac {\dot {h}}{2} = - {\pi}^4 g^2 \biggl 
(\frac {3 \delta {\rho}_x}{{m_x}^3} - \frac {2 \bar {{\rho}_q}' 
\delta {{\rho}_q}'}{{m_q}^3 \bar {\rho}_x} - \frac {\bar {{\rho}_q}'^2 
\delta {{\rho}_x}}{{m_q}^3 \bar {{\rho}_x}^2}\biggr ) \label {deltrhoxeq}
\ee
Effect of interactions in (\ref {deltrhoxeq}) is negligible and evolution 
of matter fluctuations is practically the same as the standard 
$\Lambda$CDM case. In the limit $\dot {\bar {\phiq}} \rightarrow 0$, 
we find the following relation between spatial fluctuation of 
$\delta \phiq$ and $\delta {u_x}^i$:
\be
- V' (\bar{\phiq}, \bar {\rho}_x) {\partial}^i (\delta 
\phiq) = {\Gamma}_q \bar {\rho}_x \delta {u_x}^i \label {phidotueq}
\ee
Equation (\ref {phidoteq}) shows that the divergence of quintessence field 
fluctuations ${\partial}^i \delta \phiq$ follows the velocity 
dispersion of the dark matter with opposite direction, but amplitude of 
fluctuations is largely reduced due to the very small decay width 
${\Gamma}_q$. With the expansion of the Universe, $V' (\bar{\phiq},
 \bar {\rho}_x)$ varies only very slightly - just the interaction between 
SDM and $\phiq$ will change when $\bar {\rho}_x$ decreases by a factor of 
$a^{-3}(t)$ - and even gradual increase of the dark matter clumping and 
therefore the velocity dispersion $\delta {u_x}^i$ ~\cite {sdss} can not 
eliminate the effect of decreasing density. The conclusion is that the 
spatial variation of $\phiq$ is very small from the beginning and 
practically unobservable.

\section{Production and Physics of SDM} \label{sec:prodphys}
Investigation of roles for SDM is not complete without considering 
mechanisms by which these huge particles can be produced. It is also 
inevitable that we must be able to find a proper place for them in the 
zoo of particle physics models. 

\subsection{Production} \label{sec:prod}
According to our knowledge of the early universe - which is not yet 
completely proved - SDM like all other particles should be 
produced in an epoch of preheating/reheating just at the end of inflation. 
Production of such massive particles however is not an easy task. If 
preheating/reheating has happened at scales higher than SDM mass, dangerous 
amount of gravitinos and modulies should be produced which according to 
most popular models decayed after nucleosynthesis and compromised present 
observations of primordial deuterium and $He$. The energy scale of 
preheating/reheating therefore should be lower than $\sim 10^9 GeV$.
Note that there is a 
difference between the scale of particle production and the maximum 
temperature during reheating. The boom in particle production is during 
preheating and it is not thermal i.e. particle production is too fast to 
permit a thermal equilibrium to happen. Depending on the inflation model 
after a few (or even in some cases one~\cite{instinf}) oscillation of 
inflaton at the 
bottom of its potential, most of its energy is transferred to other 
particles and interaction between them creates a thermal plasma. Not 
all species however necessarily arrive to a thermal equilibrium. Species 
with very weak interaction can decouple before getting thermalized. 

During thermalization the production of particles continues both for heavy 
species and for lighter ones. If the dominant field $\psi$ - it can be 
inflaton or another field - is heavier than $X$, after its production it 
is relativistic, otherwise it is not. Here the field $X$ can be any field, 
but we are specially interested on SDM. Production of heavy particles is 
controlled by what is 
called $T_{max}$ which depends on ${\Gamma}_{\psi}$ the decay 
width of $\psi$ and on the Hubble Constant at the beginning of reheating. 
It is the maximum temperature of the plasma before it cools due to 
expansion. Reheating temperature $T_{rh}$ depends only on ${\Gamma}_{\psi}$. 
It has been shown~\cite{infxprod} that today contribution of heavy particle $X$ which has 
been produced in a non-equilibrium, non-relativistic condition is:
\be
{\Omega}_X h^2 \approx 1 \times 10^{-5}\frac{g}{2}\biggl [\frac{g_*(T_{rh}}
{10}\biggr ]\biggl [\frac{10}{g_*(T_*)}\biggr ]^2\biggl [\frac{T_{rh}}{100 MeV}\biggr ]^5\biggl [\frac{100 GeV}{m_X}\biggr ]^4  \label{reheatcont}
\ee
where $T_*$ is the temperature at maximum particle production, $g$ and $g*$
are internal degrees of freedom respectively for $X$ and for all species.
$g_*$ depends on the particle physics at preheating/reheating scale. One 
expects that it is of order 
$100$. It is easy to see that (\ref{reheatcont}) leads to ${\Omega}_X \sim 
{\mathcal O}(0.1)$ only if $T_{rh}$ is close to dangerous limit of 
$\sim 10^9 GeV$. For lower reheating temperature $X$ can not dominate CDM 
today.

There is however another phenomenon which can produce SDM more efficiently: 
Strong gravitation at the end of inflation~\cite{infgravprod}. It has been 
shown in detail for a 
hybrid inflation. The difference between quantum vacuum when the massive 
scalar field begins to roll-down from the false vacuum to its real vacuum at 
the end of inflation appears as particle production at late time. Constraints 
from having a successful inflation with enough e-folding, etc. limits the 
mass of massive fields 
to $\lesssim 10^{-3} M_{p}$. To order of magnitude precision, the 
contribution of these heavy particles to the CDM today is (decay of SDM is 
not included):
\be
{\Omega}_X h^2 \approx \biggl (\frac{m_X}{10^{11} GeV}\biggr )^2 
\biggl (\frac{T_{rh}}{10^9 GeV}\biggr )
\ee
which for $m_X \sim 10^{13} GeV$ and $T_{rh} \sim 10^4 GeV$, 
${\Omega}_X h^2 \sim 1$. A more precise evaluation of contribution needs 
detail knowledge of parameters and a more precise calculation of quantum 
and classical phenomena. This rough estimation however is enough to show the 
possibility of having a dominant SDM.

\subsection{Particle Physics} \label{sec:phys}
It is usually assumed that highest mass scale in a field theory is less or 
around the scale of validity of the theory. SDM must have a mass 
$\gtrsim 10^{12} GeV$. Therefore if the GUT scale is $\sim 10^{16} GeV$ we 
expect to find particles of this mass range in GUT candidate 
theories~\cite {highmass}. The challenge however is to make them 
meta-stable with a lifetime greater than present age of the Universe. This 
needs either a very small coupling with high-order non-perturbative 
interactions or global symmetries similar to baryon number which are very 
softly and non-perturbatively broken.

Phenomenologically, the decay lagrangian of a field $X$ can be written as:
\bea
{\mathcal L} \sim \frac {g}{M_*^p} X {\phi}^m {\psi}^n. \label {declag} \\
p = d_x + m + \frac {3}{2} n - 4.  \label {coupdim}
\eea
where $\phi$ and $\psi$ are respectively generic bosonic and fermionic 
fields. $g$ is a dimensionless coupling constant and $M_*$ is Planck mass 
scale or any other natural mass scale in the theory. This lagrangian leads 
to a lifetime $\tau$:
\be
\tau \sim \frac {1}{g^2M_X} ({\frac {M_*}{M_X}})^{2p}. \label {lif}
\ee
For $M_X \lesssim M_*$, the exponent $p$ must be large and (\ref {declag}) 
becomes non-renormalizable. The other possibility is an extremely suppressed 
coupling constant. The latter however would not be very natural unless the 
coupling is effective and related for instance to the physics at a higher 
scale. 

High order lagrangians can be found in (SUSY)GUT models usually inspired 
by String/M-Theory~\cite{mth} (heterotic strings and quantum 
gravity in 11-dim. models). Some compactification scenarios in string 
theory predict composite particles (e.g. {\it cryptons}) with large symmetry 
groups~\cite {crypton1}~\cite {crypton2} and $M \gtrsim 10^{14} GeV$. 
The general feature of this class of models is having a very large symmetry 
group of type $G = \prod_i SU (N_i) \bigotimes \prod_j SO (2n_j)$. Their 
particle contents includes light particles with fractional charges which 
have not 
been observed. It is therefore believed that they are confined at very 
high energies $> 10^{10-12} GeV$. All of their decay modes are of 
type (\ref {declag}) and their lifetime is in the necessary range.

Models with discrete symmetries seems more natural specially because they 
have counterparts at low energies. Anomaly cancellation condition 
restricts discrete groups to ${\bf Z}_2$ and ${\bf Z}_3$~\cite{dissy}. 
A number of examples of discrete symmetries exist in Standard Model: 
Parity conservation and baryon parity which is proposed to be 
responsible for proton stability~\cite {z3}.

$SO (10)$-SUSY model is one of the favorite GUT candidates and some 
implementation of 
this model may include field with necessary characteristics of SDM. 
Messenger bosons responsible for communicating the soft SUSY breaking to 
the visible sector have masses $\gtrsim 10^{14} GeV$~\cite {highmass}. 
Messengers in representation $({\bf 8, 1})_0$ and $({\bf 1, 3})_0$ of 
Standard Model $SU (3) \bigotimes SU (2) \bigotimes U (1)$ have been 
proposed as SDM and $Y$ - a non-SM particle in which SDM decays~\cite{dissy}. 
However, in this case SDM would have strong interaction and it would be 
difficult to explain the large observed bias between dark matter and 
baryons in present universe. Moreover, in the early universe before 
nucleosynthesis, its large mass and strong interaction with quark-gluon 
plasma could create small scale anisotropies with important implications 
for galaxy formation. These perturbations have not been observed and 
in fact for explaining the distribution of galaxies today, it is 
necessary to wash out very small scale anisotropies. By contrast, 
$({\bf 1, 3})_0$ 
representation for SDM particles is a more interesting possibility because 
in this case they have only weak interaction with ordinary matter and no 
interaction with photons. This may explain some of features of galaxy 
distribution and CMB small scale anisotropies.

Other scenarios for SDM decay are suggested: decay through quantum 
gravity processes like wormhole production~\cite {xx} and through 
non-perturbative effects like instanton production~\cite {xpart}. Inspired 
by recent interest on non-compact extra-dimensions and brane models, decay 
though gravitation or other fields which propagate in the bulk has been also 
suggested as the reason for very long life of these 
particles~\cite{xpartbran}.

\section {Closing Remarks}
In this chapter we have tried to find a solution for three puzzles of today 
physics. In contrast to some other issues like SUSY, macroscopic 
extra-dimensions, existence of Higgs particles, etc. which are motivated 
by theories, observational evidence for existence of these phenomena has 
been accumulated since at least a couple of decades.

The interesting point about models proposed here is that they are all 
related to one concept: {\it The existence of a long life 
superheavy particle}. Physicists love {\it unifications}, not just for the sake of 
having an elegant model but also because nature has learned us that there is 
no isolated entity or law in the Universe. It is in fact a logical 
necessity. If there is an isolated entity, by definition it does not 
interact with other entities in the Universe and therefore it is like if 
it doesn't exist at all.

On the more practical side unification reduces the range of possibilities 
and simplifies searches. For instance if the model for dark matter and Dark 
Energy we presented here are the way Nature works, it strongly constrains 
(SUSY)GUT. Such model most have both a $X$ type massive particle and 
intimately connected to it a very light axion type field. Because of this 
close relation which probably should be due to a conserved global symmetry 
one can imagine that a sort of seesaw mechanism is responsible for such 
huge mass separation of $X$ and $\phiq$. Seesaw mechanism has been 
also suggested to relate quintessence field to neutrinos~\cite{seesaw}.

We need yet more and better observations to confirm or rule out this model.
As mentioned before observation of UHECRs anisotropy is very preliminary 
and the data volume as well as our understanding of the distribution of 
local dark matter in the halo of Milky Way and local group is vague. With 
ground and 
space based observatories like Auger, Airwatch, etc. we should better 
understand UHECRs anisotropy and whether they are more correlated to the 
halo of the Galaxy or to the nearby extra-galactic sources.

Although the present limit on the amount of the hot DM can not constrain 
SDM model, a better understanding of its contribution to the total density 
and its contents can help to understand the physics and the nature of SDM. 

Observation of $w_q$ and its cosmological evolution is crucial for any 
model of dark energy. Supernova Cosmology Project, High-z SN Project and 
specially SNAP which will increase the statistics of SN data by few orders 
of magnitude will give us the opportunity to verify the model, whether 
$w_q < -1$ 
and how far from $-1$ it is and whether SDM model can explain observations 
without fine-tuning. Observation of small anisotropy in the DE density and 
its correlation with matter anisotropy also can be used as a signature 
of relation/interaction between DM and DE.

The small coupling of $\phiq$ with SM particles suppresses the 
probability of its direct detection. However, the detection of an 
axion-like particle e.g. the QCD axion can be a positive sign for the 
possibility of existence of $\phiq$-like particles in the Nature. The 
interesting point in SDM model is that in contrast to many others, it does 
not need a very light axion. Moreover, the range of $\phiq$ mass which SDM 
model needs is roughly in the range of axion mass not yet excluded by 
experiments. 
There is therefore hopes that next generation of experiments find the QCD 
or other very light particles.

%% file: fatdm.bbl
\def\Journal#1#2#3#4{{#1} {\bf #2}, (#3) #4}

\def\AA{\em A.\& A.}
\def\APJ{\em ApJ.}
\def\APP{\em Astropart. Phys.}
\def\APS{\em ApJ.Suppl.}
\def\AST{\em Astron. J.}
\def\CAP{\em JCAP}
\def\CPC{\em Comp. Phys. Com.}
\def\CQG{\em Class.Quant.Grav.}
\def\EJP{{\em Europ. J. Phys.} C}
\def\GCS{\em Grav.Cosmol.Suppl.}
\def\GEN{Pisma Zh. Eksp. Teor.}
\def\GRG{\em Gen. Rel. Grav}
\def\HPA{\em Helv. Phys. Acta}
\def\IMA{{\em Int. J. Mod. Phys.} A}
\def\IMD{{\em Int. J. Mod. Phys.} D}
\def\IMP{\em Int. J. Mod. Phys.}
\def\JHE{\em J. High Ener. Phys.}
\def\JPG{\em J. Phys. G: Nucl. Part. Phys.}
\def\JPg{\em J. Phys. G}
\def\JPL{\em JETPhys. Lett.}
\def\LNP{\em Lect. Notes Phys.}
\def\LRR{\em Living Rev.Rel.}
\def\JPT{\em Sov. Phys. JETP}
\def\MPA{{\em Mod. Phys.} A}
\def\MPL{{\em Mod. Phys. Lett.} A}
\def\MRA{\em MNRAS}
\def\NAT{\em Nature}
\def\NCA{\em Nuovo Cimento}
\def\NPB{{\em Nucl. Phys.} B}
\def\NPP{\em Nucl. Phys.Proc.Suppl.}
\def\PAN{\em Phys.Atom.Nucl.}
\def\PLB{{\em Phys. Lett.}  B}
\def\PLO{\em Publ. Lick Obs.}
\def\PRL{\em Phys. Rev. Lett.}
\def\PRD{{\em Phys. Rev.} D}
\def\PRe{{\em Phys. Rev.} E}
\def\PRE{\em Phys. Rep.}
\def\PRV{\em Phys. Rev.}
\def\PUS{\em Phys. Usp.}
\def\PTR{{\em Phil. Trans. R.Soc. London} A}
\def\PNA{\em Proc. N.A.S.}
\def\RMP{\em Rev.Mod.Phys.}
\def\RPP{\em Rept. Prog. Phys.}
\def\SCI{\em Science}
\def\SSR{\em Space Sci. Rev.}
\def\ZPC{{\em Z. Phys.} C}
\def\ZPH{\em Z. Phys.}